\documentclass[11pt,reqno,twoside]{amsart}

\usepackage{graphicx}

\font\code=cmtt10

\def\pmb#1{\setbox0=\hbox{$#1$}%
             \kern-.027em\copy0\kern-\wd0
             \kern+.009em\copy0\kern-\wd0
             \kern+.009em\copy0\kern-\wd0
             \kern+.009em\copy0\kern-\wd0
             \kern+.009em\copy0\kern-\wd0
             \kern+.009em\copy0\kern-\wd0
             \kern+.009em\copy0\kern-\wd0
             \kern-.045em\raise+.012em\copy0\kern-\wd0
             \kern+.009em\raise+.012em\copy0\kern-\wd0
             \kern+.009em\raise+.012em\copy0\kern-\wd0
             \kern+.009em\raise-.012em\copy0\kern-\wd0
             \kern+.009em\raise-.012em\copy0\kern-\wd0
             \kern-.018em\copy0\kern-\wd0\raise-.012em\box0}
\def\Pmb#1{\setbox0=\hbox{$#1$}%
             \kern-.033em\copy0\kern-\wd0
             \kern+.011em\copy0\kern-\wd0
             \kern+.011em\copy0\kern-\wd0
             \kern+.011em\copy0\kern-\wd0
             \kern+.011em\copy0\kern-\wd0
             \kern+.011em\copy0\kern-\wd0
             \kern+.011em\copy0\kern-\wd0
             \kern-.055em\raise+.015em\copy0\kern-\wd0
             \kern+.011em\raise+.015em\copy0\kern-\wd0
             \kern+.011em\raise+.015em\copy0\kern-\wd0
             \kern+.011em\raise-.015em\copy0\kern-\wd0
             \kern+.011em\raise-.015em\copy0\kern-\wd0
             \kern-.022em\copy0\kern-\wd0\raise-.015em\box0}
\def\ms{\medskip}

\def\ub{\underbar}
\def\noi{\noindent}

\def\fvec{{\bf f}}
\def\gvec{{\bf g}}

\def\rvec{{\bf r}}
\def\svec{{\bf s}}

\def\uvec{{\bf u}}
\def\vvec{{\bf v}}

\def\xvec{{\bf x}}

\def\Dvec{{\bf D}}

\def\Mmat{{\bf M}}

\def\dt{{\Delta t}}

\def\Grd{\nabla}
\def\Div{\nabla \cdot}

\def\duvec{\Delta {\bf u}}

\vfill

\usepackage[dvips,bottom=1.4in,right=1in,top=1in, left=1in]{geometry}

	\title{DETAILED SIMULATION OF VIRAL PROPAGATION IN THE BUILT ENVIRONMENT}

\author{Rainald L\"ohner}
\address{Center for Computational Fluid Dynamics, 
            College of Science, George Mason University, 
            Fairfax, VA 22030-4444, USA.}
\email{rlohner@gmu.edu}            

\author{Harbir Antil}
\address{Center for Mathematics and Artificial Intelligence, 
            College of Science, 
            George Mason University,
            Fairfax, VA 22030-4444, USA}
\email{hantil@gmu.edu}

\author{Sergio Idelsohn}
\address{ICREA, Catalan Institution for Research and Advanced Studies 
            and 
            CIMNE, International Center for Numerical Methods in 
            Engineering, Barcelona, Spain.}            

\author{Eugenio O\~nate}
\address{CIMNE, International Center for Numerical Methods in 
            Engineering, Barcelona, Spain.}

\thanks{H. Antil is partially supported by NSF grants DMS-1818772, DMS-1913004, 
the Air Force Office of Scientific Research (AFOSR) under 
Award NO: FA9550-19-1-0036, and Department of Navy, Naval PostGraduate 
School under Award NO: N00244-20-1-0005.
}


\begin{document}

\maketitle

\begin{abstract}
A summary is given of the mechanical characteristics
of virus contaminants and the transmission via droplets and aerosols.
The ordinary and partial differential equations describing the
physics of these processes with high fidelity
are presented, as well as appropriate numerical schemes to solve them.
Several examples taken from recent evaluations of the built environment
are shown, as well as the optimal placement of sensors.
\end{abstract}

\section{The Covid-19 Crisis}
Starting in Wuhan, China, in the fall of 2019, the Covid-19 pandemic
has claimed and will continue to claim millions of infected patients
and hundreds of thousands of deaths. The lockdowns that followed its
outbreak have led to mass unemployment, stalled economic development 
and loss of productivity that will take years to recover. Some changes
in habits and lifestyles may be permanent: in the future, working from 
home or in a `socially distanced manner' may be the prevalent modus 
operandi for large segments of society. \\
The present paper gives a short description of computational techniques
that can elucidate the flow and propagation of viruses or other
contaminants in built environments in order to mitigate or avoid infections.


\section{Virus Infection}
Before addressing the requirements for the numerical simulation of
virus propagation a brief description of virus propagation and
lifetime are given. 
Covid-19 is one of many corona-viruses. The virus is usually present 
in the air or some surface, and makes its way into the body either
via inhalation (nose, mouth), ingestion (mouth) or attachment
(eyes, hands, clothes). In many cases the victim inadvertedly
touches an infected surface or viruses are deposited on its hands, 
and then the hands touch either the nose, the eyes or the mouth, thus
allowing the virus to enter the body. \\
An open question of great importance for all that will follow is
how many viruses it takes to overwhelm the body's natural defense
mechanism and trigger an infection. This number, which is sometimes
called the {\sl viral load} or the {\sl infectious dose} will depend on
numerous factors, among them the state of immune defenses of the 
individual, the timing of viral entry (all at once, piece by piece),
and the amount of hair and mucous in the nasal vessels.
In principle, a single organism in a favourable environment 
may replicate sufficiently to cause disease \cite{Tan06}.
Data from research performed on biological
warfare agents \cite{Fra97} suggests that both bacteria and
viruses can produce disease with as few as 1-100
organisms (e.g. brucellosis 10-100, Q fever 1-10,
tularaemia 10-50, smallpox 10-100, viral haemorrhagic fevers 
1-10 organisms, tuberculosis 1). Compare these numbers and consider
that as many as 3,000 organisms can be produced by talking for
5~minutes or a single cough, with sneezing producing many more 
\cite{Lou67,Lin10,Teu10,Mil13,Wei16}. 
Figure~1, reproduced from \cite{Teu10}, shows a typical number and
size distribution.

\begin{figure}[htb]
	\centering
	\includegraphics[width=10cm]{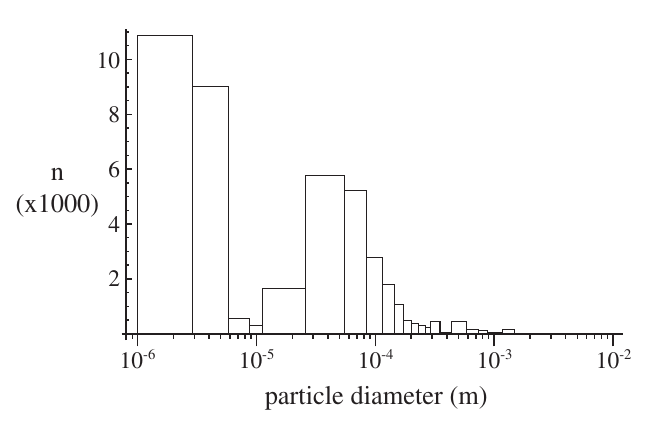}
	\caption{Counts of Particles of Various Diameters in Air Expelled by (90) Coughs \cite{Lou67}.}
\end{figure}

\section{Virus Lifetime Outside the Body}
Current evidence points to lifetimes outside the body that can range
from 1-2 hours in air to several days on particular surfaces
\cite{Kam20,vDo20}.
There has also been some documentation of lifetime variation depending
on humidity.

\section{Virus Transmission}

\subsection{Human Sneezing and Coughing}
In the sequel, we consider human sneezing and coughing as the main
conduits of virus transmission. Clearly, breathing and talking
will lead to the exhalation of air, and, consequently the exhalation
of viruses for infected victims \cite{Asa20a,Asa20b}. 
However, it stands to reason that
the size and amount of particles released -~and hence the amount of 
viruses in them~- is much higher and much more concentrated when
sneezing or coughing \cite{Fab08,Teu10,Joh11,Lin12,Asa20a,Asa20b}. \\
The velocity of air at a person's mouth during sneezing and coughing
has been a source of heated debate, particularly in the media.
The experimental evidence points to exit velocities of the order
of 2-14~m/sec \cite{Gup09,Gup10,Tan12,Tan13}. 
A typical amount and size of particles can be seen in Figure~1.

\subsection{Sink Velocities}
If, for the sake of argument, we consider Stoke's law for the drag of 
spherical particles, valid below Reynolds numbers of $Re=1$,
the terminal sink velocity (also known as the
settling velocity) of particles will be given by:

$$ v_s = {{(\rho_p - \rho_g) g \cdot d^2 } \over {18 \mu}}  ~~, \eqno(1) $$

\noi
where $\rho_p, \rho_g, g, \mu, d$ denote the density of the particles
(essentially water in the present case), density of the gas (air),
gravity, dynamic viscosity of the gas and diameter of the particle
respectively. The equivalent Reynolds' number is:

$$ Re = {{\rho_g v_s d } \over {\mu}} 
      = {{\rho_g (\rho_p - \rho_g) g d^3 } \over {18 \mu^2}} \eqno(2) $$

\noi
With $\rho_p=1~gr/cc, \rho_g=0.0012~gr/cc, 
g=981~cm/sec^2, \mu=1.81 \cdot 10^{-4}~gr/(cm \cdot sec)$ this 
yields a limiting diameter for $Re=1$ of

$$ d_{Re=1} = 0.0079~cm ~~, \eqno(3) $$

\noi
i.e. approximately 1/10th of a millimeter in diameter -~a particle
size that would still be perceived by the human eye. The corresponding
sink velocity is given by:

$$ v_s = 3 \cdot 10^{5} d^2 cm/sec ~~, \eqno(4) $$

\noi
with $d$ in $cm$, i.e. for $Re=1$

$$ v_s(Re=1) = 18~cm/sec ~~. \eqno(5) $$

\noi
Note the quadratic dependency of the sink velocity with diameter.
Table~1 lists the sink velocities for water droplets of different
diameters in air. One can see that below diameters of $O(0.1~mm)$ the
sink velocity is very low, implying that these particles remain in
and move with the air for considerable time (and possibly distances).

\begin{table}[htbp]
\begin{center}
\caption{Sink Velocities and Reynolds Number For Water Particles in Air}
\label{tab:Sinkvelo}
\begin{tabular}{c|c|c}
\hline
Diameter [mm] & sink velocity [m/sec] & Re \\
\hline
1.00E-00      & 3.01E+01      & 1.99E+03 \\
1.00E-01      & 3.01E-01      & 1.99E+00 \\
1.00E-02      & 3.01E-03      & 1.99E-03 \\
1.00E-03      & 3.01E-05      & 1.99E-06 \\
1.00E-04      & 3.01E-07      & 1.99E-09 \\
\hline
\end{tabular}
\end{center}
\end{table}

\subsection{Evaporation}
Depending on the relative humidity and the temperature of the
ambient air, the smaller particles can evaporate in milliseconds.
However, as the mucous and saliva evaporate, they build a gel-like
structure that surrounds the virus, allowing it to survive. This implies
that extremely small particles with possible viruses will remain 
infectious for extended periods of times - up to an hour according to
some studies. \\
An important question is then whether a particle/droplet will
first reach the ground or evaporate. Figure~2, taken from \cite{Xie07},
shows that below 120~$\mu m$ the particles evaporate before
falling 2~m (i.e. reaching the ground). 

\begin{figure}[h!]
	\centering
	\includegraphics[width=10cm]{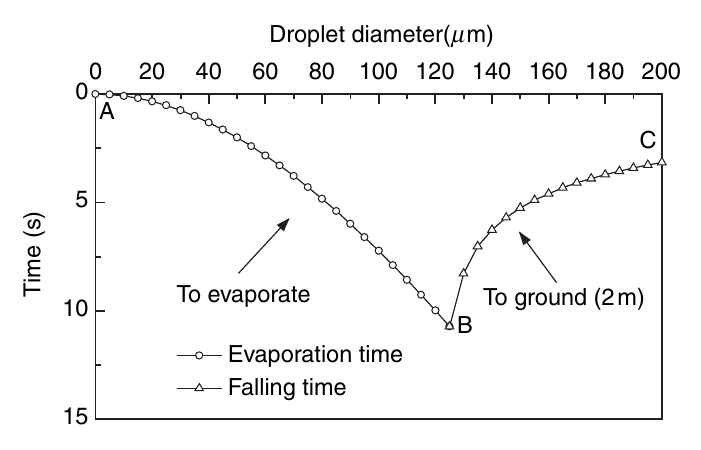}
	\caption{Evaporation Time and Falling Time of Droplets of Varying Diameter
		     ($T_{p0}=33^oC, T_{\infty}=18^oC, RH=0\%$) From \cite{Xie07}}
\end{figure}

\subsection{Viral Load}
A central question that requires an answer is then: 
how many viruses are in these small 
particles~? An approximate answer may be obtained from the experiments
that are being carried out on animals to trace and monitor infections.
For ferrets \cite{Kim20} $O(10^5-10^6)$ have been used to infect via
intranasal swabs, while for mice \cite{To20} $O(10^4)$ seem to suffice.
Viral titers can vary a lot, but one may assume 
on the order of $O(10^6)$ viruses/ml for a nasopharyngeal swab
\cite{Kim20,To20}. Table~2 lists the number of viruses per droplet and
the number of droplets needed to contain just 1~virus. Note that 
while for a droplet with a diameter of 1~mm one can expect $O(500)$
viruses, only every 2,000th particle of diameter 10~$\mu m$ does
contain a single virus.

\begin{table}[htbp]
\begin{center}
\caption{Estimated Number of Viruses For Different Particle Diameters}
\label{tab:Viralload}
\begin{tabular}{c|c|c|c}
\hline
Droplet Diameter [mm] & Volume [mm$^3$] & Viruses/droplet 
                      & Droplets Needed for 1 Virus \\
\hline
1.00E+00 & 5.24E-01 & 5.24E+02 & 1.00E+00 \\
1.00E-01 & 5.24E-04 & 5.24E-01 & 1.91E+00 \\
1.00E-02 & 5.24E-07 & 5.24E-04 & 1.91E+03 \\
1.00E-03 & 5.24E-10 & 5.24E-07 & 1.91E+06 \\
\hline
\end{tabular}
\end{center}
\end{table}

\noi
Similar numbers are seen in field studies as well.
The size of viruses varies from 0.02-0.3 $\mu m$, while the size 
of bacteria varies from 0.5-10 $\mu m$.
The influenza virus RNA detected by quantitative polymerase chain 
reaction in human exhaled breath suggests that it may
be contained in fine particles generated during tidal breathing
and not only coughs \cite{Fab08,Lin10,Lin12,Mil13}. 
Influenza RNA and Mycobacterium tuberculosis have been reported in
particles that range in size from 0.5-4.0 $\mu m$ 
(\cite{Fab08,Lin10,Lin12,Mil13} and references cited therein).

\section{Physical Modeling of Aerosol Propagation}
\label{physmodelaerosols}
When solving the two-phase equations, the air, as a continuum,
is best represented by a set of partial differential equations 
(the Navier-Stokes equations) that are numerically solved on a mesh. 
Thus, the gas characteristics are calculated 
at the mesh points within the flowfield. 
However, as the droplets/particles may be
relatively sparse in the flowfield, they can be modeled by either:
\begin{itemize}
\item[a)] A continuum description, i.e. in the same manner as the fluid
flow, or
\item[b)] A particle (or Lagrangian) description, where individual 
particles (or groups of particles) are monitored and tracked in the 
flow.
\end{itemize}
\par \noi
Although the continuum (so-called multi-fluid) method has been
used for some applications, the inherent assumptions of the continuum 
approach lead to several disadvantages which may be countered
with a Lagrangian treatment for dilute flows.
The continuum assumption cannot robustly account
for local differences in particle characteristics, particularly if the
particles are polydispersed. In addition, the only boundary conditions 
that can be considered in a straightforward manner are slipping and 
sticking, whereas reflection boundary conditions, such as specular and
diffuse reflection, may be additionally considered with a Lagrangian 
approach. Numerical diffusion of the particle density is
eliminated by employing Lagrangian particles due to their pointwise 
spatial accuracy. \\
While a Lagrangian approach offers many potential advantages, this 
method also creates problems that need to be addressed. For
instance, large numbers of particles may cause a Lagrangian analysis 
to be memory intensive. This problem is circumvented by treating 
parcels of particles, i.e. doing the detailed analysis for one 
particle and then applying the effect to many. In addition, continuous 
mapping and remapping of particles to their respective elements may 
increase computational requirements, particularly for unstructured 
grids. 

\subsection{Equations Describing the Motion of the Air}
As seen from the experimental evidence, the velocities of air
encountered during coughing and sneezing never exceed a Mach-number
of $Ma=0.1$. Therefore, the air may be assumed as a Newtonian,
incompressible liquid, where buoyancy effects are modeled via
the Boussinesq approximation. The equations describing the
conservation of momentum, mass and energy for incompressible, 
Newtonian flows may be written as

$$ \rho \vvec_{,t} + \rho \vvec \cdot \Grd \vvec + \Grd p = 
   \nabla \cdot \mu \nabla \vvec + \rho \gvec + \beta \rho \gvec(T - T_0) 
      + \svec_v   ~~, 
                                                             \eqno(6.1) $$

$$                             \Div \vvec = 0      ~~.       \eqno(6.2) $$
$$ \rho c_p T_{,t} + \rho c_p \vvec \cdot \Grd T = 
       \nabla \cdot k \nabla T + s_e ~~, 
                                                             \eqno(6.3) $$

\noi
Here $\rho, \vvec, p, \mu, \gvec, \beta, T, T_0, c_p, k$ denote 
the density, velocity vector, pressure, viscosity, gravity vector, 
coefficient of thermal expansion, temperature, reference temperature,
specific heat coefficient and conductivity respectively, and
$\svec_v, s_e$ momentum and energy source terms (e.g. due to particles
or external forces/heat sources).
For turbulent flows both the viscosity and the conductivity are
obtained either from additional equations or directly via a
large eddy simulation (LES) assumption through monotonicity induced
LES (MILES) \cite{Bor92,Fur99,Gri02,Ide19}.

\subsection{Equations Describing the Motion of Particles/Droplets}
\label{partmotion}
In order to describe the interaction of particles/droplets with the 
flow, the mass, forces and energy/work exchanged between the 
flowfield and the particles must be defined.
As before, we denote for {\bf fluid (air)}
by $\rho, p, T, k, v_i, \mu$ and $c_p$ the density, pressure,
temperature, conductivity, velocity in direction $x_i$,
viscosity, and the specific heat at constant pressure.
For the {\bf particles}, we denote by 
$\rho_p, T_p, v_{pi}, d, c_{pp}$ and 
$Q$ the density, temperature, velocity in direction $x_i$, 
equivalent diameter, and heat transferred per unit volume. 
In what follows, we will refer to droplet and particles,
collectively as particles. \\
Making the classical assumptions that the particles may be represented
by an equivalent sphere of diameter $d$, the drag forces $\Dvec$
acting on the particles will be due to the difference of fluid and 
particle velocity:

$$ \Dvec = {{\pi d^2} \over 4} \cdot c_D \cdot
         { 1 \over 2} \rho | \vvec - \vvec_p | ( \vvec - \vvec_p ) 
                                                  ~~.  \eqno(7)       $$

\noi
The {\bf drag coefficient} $c_D$ is obtained empirically from the
Reynolds-number $Re$:

$$ Re = {{\rho | \vvec - \vvec_p | d } \over { \mu }}   \eqno(8)      $$

\noi
as (see, e.g. \cite{Sch79}):

$$ c_D = max\left(0.1 ,
{24 \over Re} \left( 1 + 0.15 Re^{0.687} \right) \right)  \eqno(9) $$

\noi
The lower bound of $c_D=0.1$ is required to obtain the proper limit for
the Euler equations, when $Re \rightarrow \infty$.
\noi
The heat transferred between the particles and the fluid is given by

$$ Q = {{\pi d^2} \over 4} \cdot 
      \left[ h_f      \cdot ( T   - T_p   )
           + \sigma^* \cdot ( T^4 - T_p^4 ) \right]
                                                        ~~,  \eqno(10) $$
\noi
where $h_f$ is the film coefficient and $\sigma^*$ the radiation 
coefficient. For the class of problems considered here, the particle 
temperature and kinetic energy are such that the radiation 
coefficient $\sigma^*$ may be ignored. The film coefficient $h_f$ is 
obtained from the Nusselt-Number $Nu$:

$$ Nu = 2 + 0.459 Pr^{0.333} Re^{0.55} ~~,  \eqno(11)        $$

\noi
where $Pr$ is the Prandtl-number of the gas

$$ Pr = {k \over \mu } ~~,  \eqno(12)         $$

\noi
as

$$ h_f = {{ Nu \cdot k }\over d} ~~.      \eqno(13)    $$

\par \noi
Having established the forces and heat flux, the particle motion 
and temperature are obtained from Newton's law and the first law 
of thermodynamics. For the particle velocities, we have:

$$ \rho_p {{\pi d^3} \over 6 } \cdot {{ d\vvec_p} \over {dt}} = \Dvec 
   + \rho_p {{\pi d^3} \over 6 } \gvec ~~.
                                                        \eqno(14)    $$

\noi
This implies that:

$$ {{ d\vvec_p} \over {dt}} = {{3 \rho} \over {4 \rho_p d}} \cdot c_d
                               | \vvec - \vvec_p | ( \vvec - \vvec_p ) 
                    + \gvec
                    = \alpha_v | \vvec - \vvec_p | ( \vvec - \vvec_p )
                    + \gvec  ~~.
                                                         \eqno(15)    $$

\noi
where $\alpha_v=3\rho c_d / (4 \rho_p d)$.
The particle positions are obtained from:

$$ {{ d\xvec_p} \over {dt}} = \vvec_p ~~.     \eqno(16)    $$

\noi
The temperature change in a particle is given by:

$$ \rho_p c_{pp} {{\pi d^3} \over 6 } \cdot {{ dT_p} \over {dt}} = Q ~~,
                                                         \eqno(17)    $$

\noi
which may be expressed as:

$$ {{ dT_p} \over {dt}} = {{3 k}\over{4 c_{pp} \rho_p d^2}} \cdot Nu \cdot
                          ( T - T_p )
                        = \alpha_T ( T - T_p ) ~~,     \eqno(18)    $$

\noi
with $\alpha_T=3 k/(4 c_{pp} \rho_p d^2)$.
Equations (15, 16, 18) may be formulated as a system of Ordinary 
Differential Equations (ODEs) of the form:

$$ {{d\uvec_p} \over {dt}} = \rvec(\uvec_p, \xvec, \uvec_f) ~~,
                                                         \eqno(19)    $$

\noi
where $\uvec_p, \xvec, \uvec_f$ denote the particle unknowns, the
position of the particle and the fluid unknowns at the position of
the particle.

\subsection{Numerical Integration of the Motion of the Air}
\label{numintnavto}
The last six decades have seen a large number of schemes that may be
used to solve numerically the incompressible Navier-Stokes
equations given by Eqns.(6.1-6.3). In the present case, the following
design criteria were implemented:
\begin{itemize}
\item[-] Spatial discretization using {\bf unstructured grids} 
(in order to allow for arbitrary geometries and adaptive refinement);
\item[-] Spatial approximation of unknowns with 
{\bf simple linear finite elements} (in order to have a simple 
input/output and code structure);
\item[-] Edge-based data structures (for reduced access to memory and
indirect addressing);
\item[-] Temporal approximation using {\bf implicit integration of viscous
terms and pressure} (the interesting scales are the ones associated with
advection);
\item[-] Temporal approximation using {\bf explicit, high-order 
integration of advective terms};
\item[-] {\bf Low-storage, iterative solvers} for the resulting systems of
equations (in order to solve large 3-D problems); and
\item[-] Steady results that are {\bf independent from the timestep} chosen
(in order to have confidence in convergence studies).
\end{itemize} 
\noindent
The resulting discretization in time is given by the following projection
scheme \cite{Loh06,Loh08}:
\begin{itemize}
\item[-] \ub{Advective-Diffusive Prediction}:
$\vvec^n, p^n \rightarrow \vvec^{*}$

$$ \svec' = - \Grd p^n + \rho \gvec 
          + \beta \rho \gvec (T^n - T_0) + \svec_v ~~,
\eqno(20)
$$

$$
\vvec^i = \vvec^n + \alpha^i \gamma \dt \left(
 - \vvec^{i-1} \cdot \Grd \vvec^{i-1} 
   \nabla \cdot \mu \nabla \vvec^{i-1} + \svec' \right)  ~~; ~~i=1,k-1~~;
\eqno(21a)
$$

$$
 \left[ { 1 \over \dt} - \theta \nabla \cdot \mu \nabla \right]
   \left( \vvec^{k} - \vvec^n \right)
 + \vvec^{k-1} \cdot \Grd \vvec^{k-1} = 
   \nabla \cdot \mu \nabla \vvec^{k-1} + \svec' ~~.  \eqno(21b)
$$

\ms \noi
\item[-] \ub{Pressure Correction}: $p^n \rightarrow p^{n+1}$

$$
 \Div \vvec^{n+1} = 0                          ~~;
\eqno(22)
$$
$$
 {{ \vvec^{n+1} - \vvec^{*} }\over \dt} + \Grd ( p^{n+1} - p^n )
   = 0                                           ~~;
\eqno(23)
$$

\noi
\item[ ] which results in

$$
 \nabla^2 ( p^{n+1} - p^n ) = {{\Div \vvec^{*} }\over \dt} ~~;
\eqno(24)
$$

\ms \noi
\item[-] \ub{Velocity Correction}:
$\vvec^{*} \rightarrow \vvec^{n+1}$

$$
 \vvec^{n+1} = \vvec^{*} - \dt \Grd ( p^{n+1} - p^n ) ~~.
\eqno(25)
$$
\end{itemize}

\noi
$\theta$ denotes the implicitness-factor for the viscous
terms ($\theta=1$: 1st order, fully implicit, $\theta=0.5$: 2nd order,
Crank-Nicholson). 
$\alpha^i$ are the standard low-storage Runge-Kutta coefficients
$\alpha^i=1/(k+1-i)$. The $k-1$ stages of Eqn. (21a) may be seen as a 
predictor (or replacement)
of $\vvec^n$ by $\vvec^{k-1}$. The original right-hand side has not been
modified, so that at steady-state $\vvec^n=\vvec^{k-1}$, preserving the
requirement that the steady-state be independent of the timestep $\dt$.
The factor $\gamma$ denotes the local ratio of the stability limit for
explicit timestepping for the viscous terms versus the timestep chosen.
Given that the advective and viscous timestep limits are proportional to:

$$ \dt_a \approx {h \over {|\vvec|}} ~~;~~
   \dt_v \approx {{\rho h^2} \over \mu} ~~, \eqno(26)
$$

\noi
we immediately obtain

$$ \gamma = {{\dt_v} \over {\dt_a}}
    \approx {{\rho |\vvec| h }\over{\mu}} \approx Re_h  ~~,
\eqno(27)
$$

\noi
or, in its final form:

$$ \gamma = min(1,Re_h) ~~. \eqno(28) $$

\noi
In regions away from boundary layers, this factor is $O(1)$, implying 
that a high-order Runge-Kutta scheme is recovered. Conversely, for 
regions where $Re_h=O(0)$, the scheme reverts back to the usual
1-stage Crank-Nicholson scheme. 
Besides higher accuracy, an important benefit of explicit multistage 
advection schemes is the larger timestep one can employ. The increase in 
allowable timestep is roughly proportional to the number of stages used 
(and has been exploited extensively for compressible flow simulations 
\cite{Jam81}). 
Given that for an incompressible solver of the projection type 
given by Eqns.(20-25) most of the
CPU time is spent solving the pressure-Poisson system Eqn.(24), the speedup
achieved is also roughly proportional to the number of stages used. \\
At steady state, $\vvec^{*}=\vvec^n=\vvec^{n+1}$ and the residuals of the
pressure correction vanish,
implying that the result does not depend on the timestep $\dt$. \\
The spatial discretization of these equations is carried out via 
linear finite elements. The
resulting matrix system is re-written as an edge-based solver, allowing
the use of consistent numerical fluxes to stabilize the advection and
divergence operators \cite{Loh08}. \\
The energy (temperature) equation (Eqn.(6.3)) is integrated in a manner
similar to the advective-diffusive prediction (Eqn(21)), i.e. with
an explicit, high order Runge-Kutta scheme for the advective parts and
an implicit, 2nd order Crank-Nicholson scheme for the conductivity.

\subsection{Numerical Integration of the Motion of Particles/Droplets}
\label{numintpast}
The equations describing the position, velocity and temperature of a
particle (Eqns.\ 15-19) may be formulated as a system of nonlinear 
Ordinary Differential Equations of the form:

$$ {{d\uvec_p} \over {dt}} = \rvec(\uvec_p, \xvec, \uvec_f) ~~.
                                                         \eqno(29)    $$

\par \noi
They can be integrated numerically in a variety of ways. Due to its
speed, low memory requirements and simplicity, we have chosen
the following k-step low-storage Runge-Kutta procedure to integrate them:

$$ \uvec^{n+i}_p = \uvec^n_p + \alpha^i \Delta t \cdot
   \rvec(\uvec^{n+i-1}_p, \xvec^{n+i-1}, \uvec^{n+i-1}_f) ~~,
~~ i=1,k  ~~. \eqno(30) $$

\noi
For linear ODEs the choice

$$ \alpha^i= {1 \over {k+1-i}} ~~,~~ i=1,k  \eqno(31) $$

\noi
leads to a scheme that is $k$-th order accurate in time.
Note that in each step the location of the particle with respect to the
fluid mesh needs to be updated in order to obtain the proper values for
the fluid unknowns. The default number of stages used is $k=4$. This
would seem unnecessarily high, given that the flow solver is of
second-order accuracy, and that the particles are integrated separately
from the flow solver before the next (flow) timestep, i.e. in a staggered
manner. However, it was found that the 4-stage particle integration
preserves very well the motion in vortical structures and leads to less
`wall sliding' close to the boundaries of the domain \cite{Loh14}.
The stability/ accuracy of the particle integrator should not be a problem
as the particle motion will always be slower than the maximum wave speed
of the fluid (fluid velocity). \\
The transfer of forces and heat flux between the fluid and the particles
must be accomplished in a conservative way, i.e. whatever is added to the
fluid must be subtracted from the particles and vice-versa. The finite
element discretization of the fluid equations will lead to
a system of ODE's of the form:

$$ \Mmat \duvec = \rvec ~~,    \eqno(32)       $$

\noi
where $\Mmat, \duvec$ and $\rvec$ denote, respectively, the consistent 
mass matrix, increment of the unknowns vector and right-hand side vector. 
Given the `host element' of each particle, i.e. the fluid mesh element 
that contains the particle, the forces and heat transferred 
to $\rvec$ are added as follows:

$$ \rvec^i_D = \sum_{el~surr~i} N^i(\xvec_p) \Dvec_p ~~.  \eqno(33)  $$

\noi
Here $N^i(\xvec_p)$ denotes the shape-function values of the host 
element for the point coordinates $\xvec_p$, and the sum extends
over all elements that surround node $i$. As the sum of all 
shape-function values is unity at every point:

$$ \sum N^i(\xvec) = 1 ~~\forall \xvec ~~,    \eqno(34)       $$

\noi
this procedure is strictly conservative. \\
From Eqns.(15-18) and their equivalent numerical integration via
Eqn.(30), 
the change in momentum and energy for one particle is given by:

$$ \fvec_p =  \rho_p {{\pi d^3}\over 6} 
             {{\left( \vvec^{n+1}_p - \vvec^n_p \right)} \over {\Delta t}}
                                           ~~,    \eqno(35)       $$

$$ q_p =  \rho_p c_{pp} {{\pi d^3}\over 6} 
             {{\left( T^{n+1}_p - T^n_p \right)} \over {\Delta t}}
                                           ~~.    \eqno(36)       $$

\noi
These quantities are multiplied by the number of particles in
a packet in order to obtain the final values transmitted to the fluid.
Before going on, we summarize the basic steps required in order to update
the particles one timestep:
\begin{itemize}
\item[-] Initialize Fluid Source-Terms: $\rvec=0$
\item[-] {\code DO}: For Each Particle:
\item[ ] - {\code DO}: For Each Runge-Kutta Stage:
\item[ ] ~~~- Find Host Element of Particle: {\code IELEM}, $N^i(\xvec)$
\item[ ] ~~~- Obtain Fluid Variables Required
\item[ ] ~~~- Update Particle: Velocities, Position, Temperature, ...
\item[-] - {\code ENDDO}
\item[ ] - Transfer Loads to Element Nodes
\item[-] {\code ENDDO}
\end{itemize}

\subsubsection{Particle Parcels}
\label{partparc}
For a large number of very small particles, it becomes impossible to
carry every individual particle in a simulation. The solution is to:
\begin{itemize}
\item[a)] Agglomerate the particles into so-called packets of $N_p$ 
particles;
\item[b)] Integrate the governing equations for one individual particle; 
and
\item[c)] Transfer back to the fluid $N_p$ times the effect of one 
particle.
\end{itemize}
Beyond a reasonable number of particles per element (typically $> 8$),
this procedure produces accurate results without any deterioration in
physical fidelity.

\subsubsection{Other Numerics}
In order to achieve a robust particle integrator, a number of additional
precautions and algorithms need to be implemented. The most
important of these are:
\begin{itemize}
\item[-] Agglomeration/Subdivision of Particle Parcels:
As the fluid mesh may be adaptively refined and coarsened in time,
or the particle traverses elements of different sizes,
it may be important to adapt the parcel concentrations as well. 
This is necessary to ensure that there is sufficient parcel 
representation in each element and yet, that there are not too many 
parcels as to constitute an inefficient use of CPU and memory. 
\item[-] Limiting During Particle Updates:
As the particles are integrated independently from the flow solver, it is
not difficult to envision situations where for the extreme cases of
very light or very heavy particles physically meaningless or unstable
results may be obtained.
In order to prevent this, the changes in 
particle velocities and temperatures are limited in order not to exceed 
the differences in velocities and temperature between the particles and 
the fluid \cite{Loh14}. 
\item[-] Particle Contact/Merging:
In some situations, particles may collide or merge in a certain region
of space.
\item[- ] Particle Tracking:
A common feature of all particle-grid applications is that the particles
do not move far between timesteps. This makes physical sense:
if a particle jumped ten gridpoints during one timestep, it would have
no chance to exchange information with the points along the way, leading
to serious errors. Therefore, the assumption that the new host elements
of the particles are in the vicinity of the current ones is a valid one.
For this reason, the most efficient way to search for the new host elements
is via the vectorized neighbour-to-neighbour algorithm described in
\cite{Loh90,Loh08}. 
\end{itemize}


\section{Examples}
\label{examples}
The techniques described above were implemented in FEFLO,
a general-purpose computational fluid dynamics (CFD) code based 
on the following general principles:
\begin{itemize}
\item[-] Use of unstructured grids (automatic grid generation and mesh
refinement);
\item[-] Finite element discretization of space;
\item[-] Separate flow modules for compressible and incompressible flows;
\item[-] Edge-based data structures for speed;
\item[-] Optimal data structures for different architectures;
\item[-] Bottom-up coding from the subroutine level to assure an
open-ended, expandable architecture.
\end{itemize}
\par \noi
The code has had a long history of relevant applications involving
compressible flow simulations in the areas of
transonic flow {\cite{Luo94,Luo94a,Luo99,Luo00,Loh01,Luo01}},
store separation {\cite{Bau93,Bau94,Bau95a,Bau97,Bau97a}},
blast-structure interaction {\cite{Bau91,Bau93a,Bau95,Bau96,Bau99},
\cite{Loh04a,Loh08a,Ric08,Tog09,Stu10}},
incompressible flows {\cite{Ram93,Ram99,Loh04,Loh06,Aub08,Loh11c}},
free-surface hydrodynamics {\cite{Loh98c,Loh06a,Loh07b}},
dispersion {\cite{Cam04,Cam04a,Loh05,Cam06}},
patient-based haemodynamics {\cite{Ceb01,Loh01,Ceb05,App08,Loh08b}} and
aeroacoustics {\cite{Liu09}}. The code
has been ported to vector {\cite{Loh02}},
shared memory {\cite{Loh98,Sha00}},
distributed memory {\cite{Ram93,Loh95,Ram96,Loh11a}} and GPU-based
\cite{Cor10a,Cor10b,Cor11a,Cor11b,Loh11b}
machines.

\medskip \noindent
The cases shown all simulate sneezing/coughing in different environments.
The ambient temperature was assumed to be $20^o$C. 
In order to simulate a sneeze/cough, the 
velocity and temperature in a spherical region of radius
($r=5~cm$) near the patient's mouth was reset 
at the beginning of each timestep according to the following
triangular function:

$$ f(t)= 
   \left\{
   \begin{array}{ll}
     {t \over t_{mid}} \hfill  & ~~\text{if}:         0 \le t \le   t_{mid} \\
    1 - {{t-t_{mid}}\over t_{mid}} \hfill & ~~\text{if}:   t_{mid} \le t \le 2 t_{mid} \\
     0                  \hfill & ~~\text{if}: 2 t_{mid} \le t             
     \end{array}
     \right. 
	~~. \eqno(37) $$

$$ v(t) = 5 f(t) ~[m/sec] ~~,~~ T(t) = 20 + f(t) ( 37 - 20 ) ~~.
    \eqno(38) $$

\noi
The droplets were initialized with 4 different sizes and different
velocities, and released every $0.005$ seconds during $0.1$ seconds.
This resulted in a final number of particle packets of $n_p=25,662$.
The temperature was set to $T_p=37^oC$ and the velocity to $v=5~m/sec$.
Table 3 summarizes the diameters and resulting mass distribution.

\begin{table}[htbp]
\begin{center}
\caption{Initial Conditions for Particles}
\label{tab:inicondparts}
\begin{tabular}{c|c|c|c}
\hline
Droplet Diameter [mm] & Mass [gr$^3$] 
                      & Nr. of Packets & Nr. of Particles \\
\hline
1.00E+00 & 5.50E+00 & 1.05E+03 & 1.05E+04 \\
1.00E-01 & 0.11E+00 & 2.10E+03 & 2.10E+05 \\
1.00E-02 & 0.58E-02 & 1.12E+04 & 1.12E+07 \\
1.00E-03 & 0.58E-05 & 1.12E+04 & 1.12E+08 \\
\hline
\end{tabular}
\end{center}
\end{table}

\noindent
In the cases shown different temporal scales appear: 
\begin{itemize}
\item[-] The fast, ballistic drop of the larger ($d=1~mm$) particles, 
occurring in the range of $O(1)~sec$;
\item[-] The slower drop of particles of diameter $d=O(0.1)~mm$, 
occurring in the range of $O(10)~sec$; and
\item[-] The transport of the even smaller particles through the air,
occurring in the range of $O(100)~sec$.
\end{itemize}
We have attempted to show these phases in the results, and for this 
reason the results are not displayed at equal time intervals.
Unless otherwise noted, the particles have been colored according to
the {\bf logarithm} of the diameter, with red colors representing the
largest and blue the smallest particles. \\
The examples given show clearly the dangers of droplet- and aerosol-
based infections in the built environment.

\subsection{Sneezing in Transportation Security Agency (TSA) Queues}
One of the obvious vectors for viral contamination and spread are
security and passport examination queues in airports. Air flow 
is moderate, passengers are in very close proximity, and in some 
airports queues wind back and forth in narrow lanes. Figure~3,4 
show the arrangement of pedestrians, as well as the discretization
chosen. Note the smaller elements close to the bodies and in the
region of interest between the two pedestrians in the middle row.
This particular mesh had {\code 12.74Mels}.
The distribution of particles and the absolute value of the velocity
in the centerplane over time can be discerned from 
Figures~5-7. One can see that the large (red) particles follow a ballistic
path and have some influence on the flow (e.g. at time $t=0.20$).
This `ballistic phase' ends at about $t=1~sec$. The (green) particles of size
$d=0.1~mm$ are quickly stopped by the air, and then sink slowly towards
the floor in close proximity to the individual sneezing. The even smaller
(cyan, blue) particles rise with the cloud of warmer air exhaled by the
sneezing individual, and disperse much further at later times.

\begin{figure}[h!]
\centering
	\includegraphics[width=10cm]{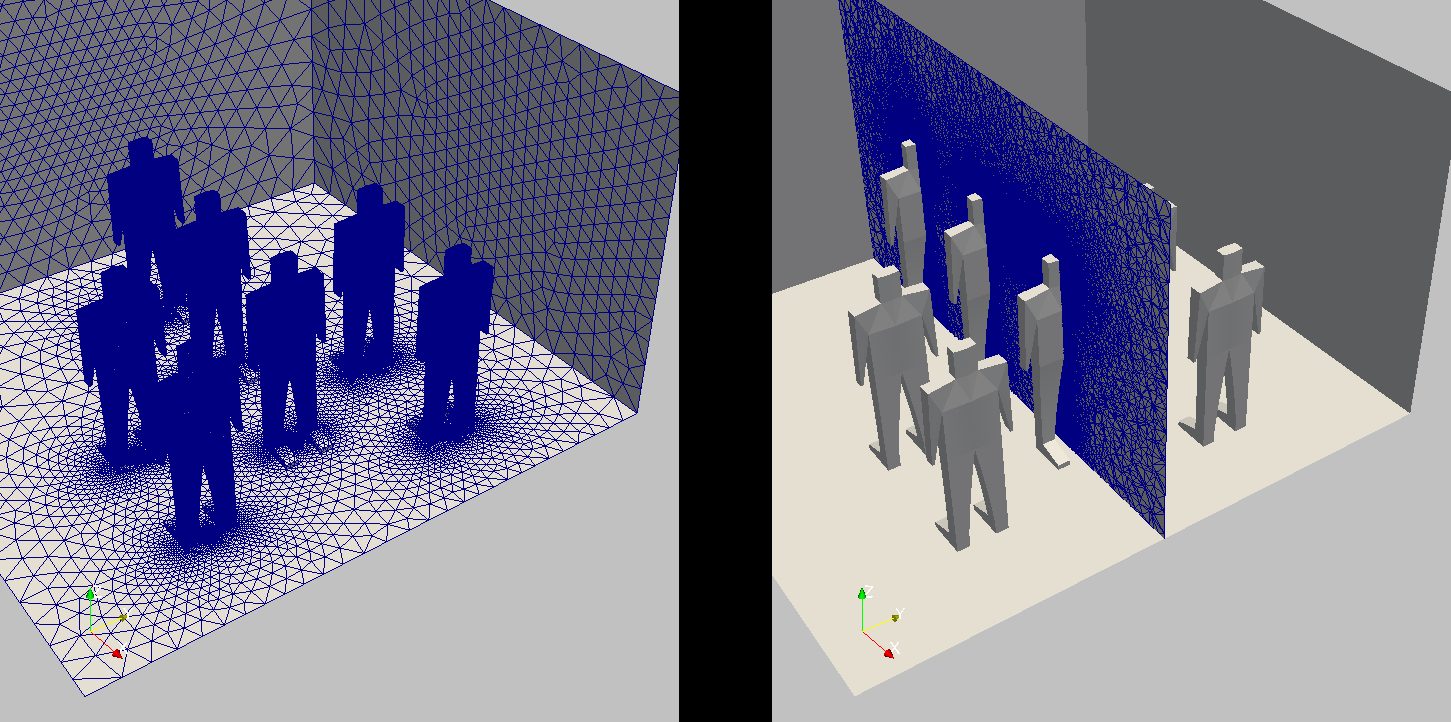}
	\caption{TSA Queue: Arrangement of Pedestrians and Surface Mesh}
\end{figure}

\begin{figure}[h!]
\centering
	\includegraphics[width=10cm]{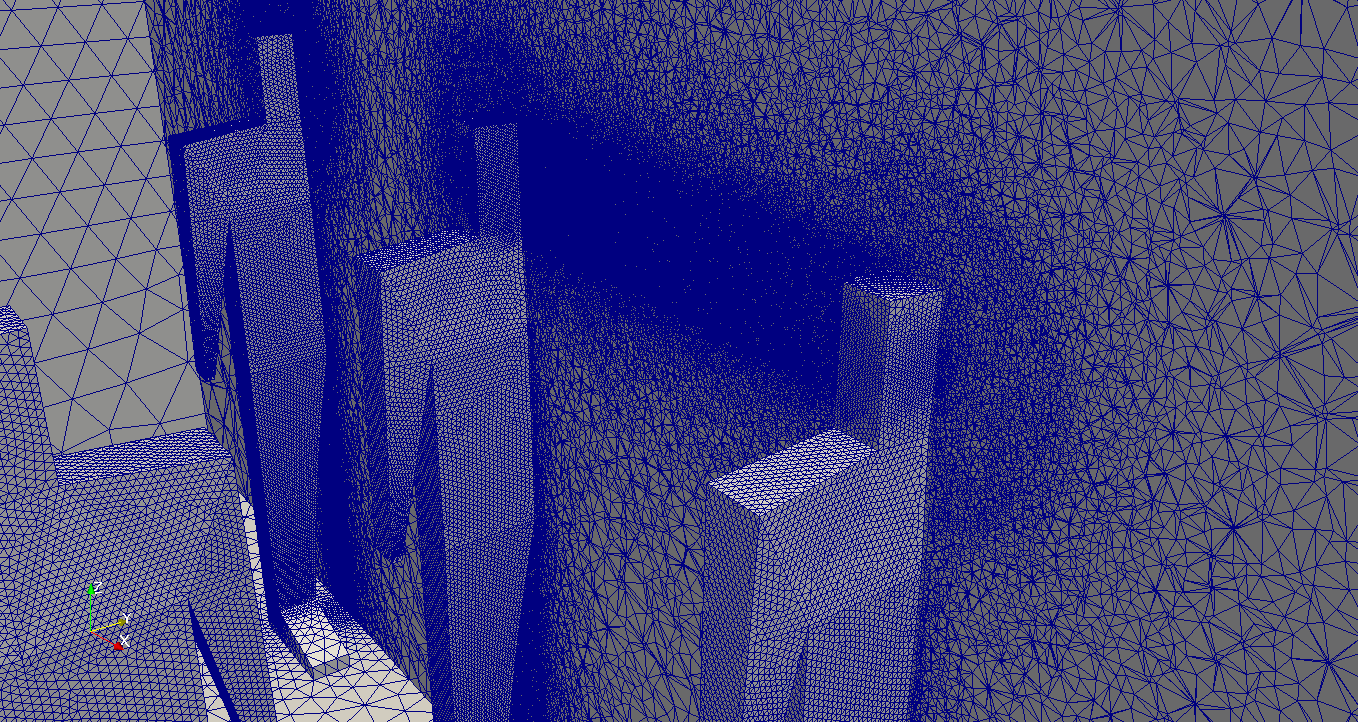}
	\caption{TSA Queue: Surface Mesh and Cut Plane}
\end{figure}

\begin{figure}[h!]
\centering
	\includegraphics[width=10cm]{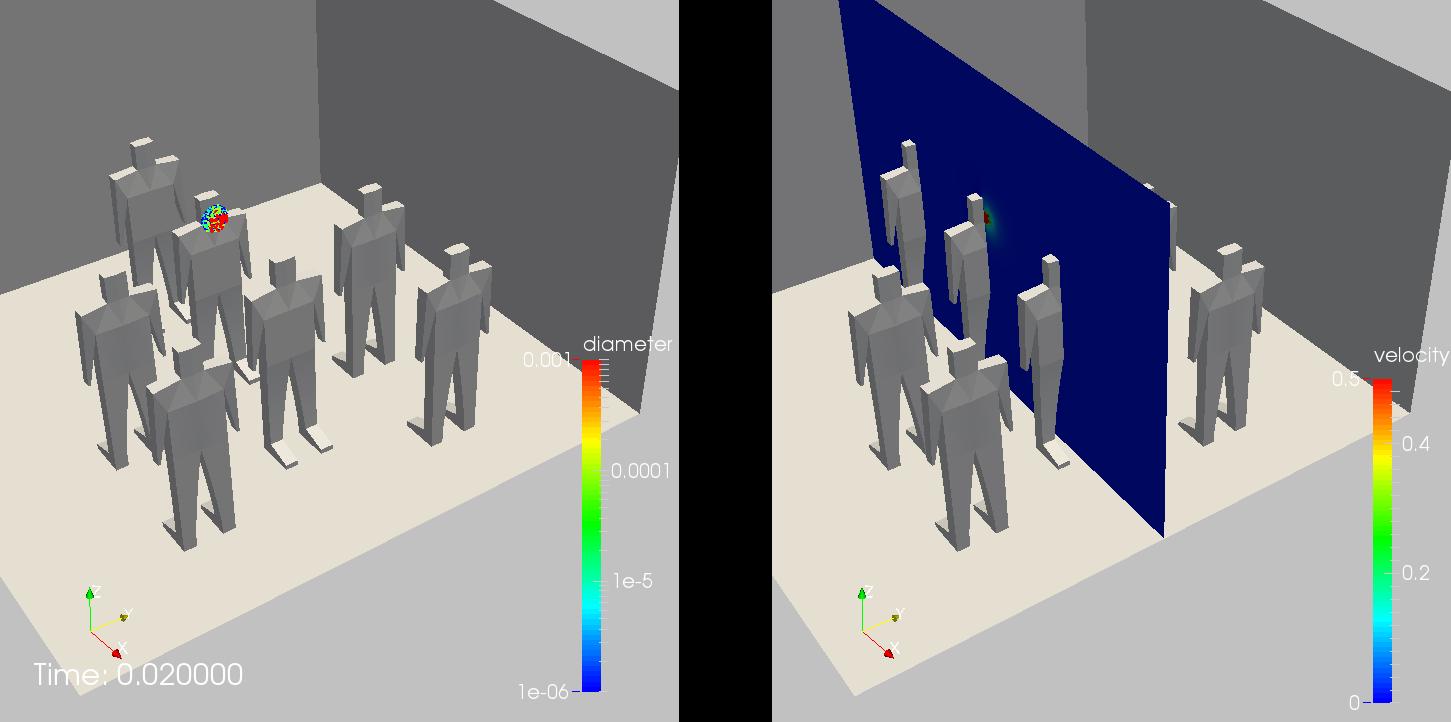}
	\caption{TSA Queue: Particle Distribution at $t=0.02~sec$}
\end{figure}

\begin{figure}[h!]
\centering
	\includegraphics[width=10cm]{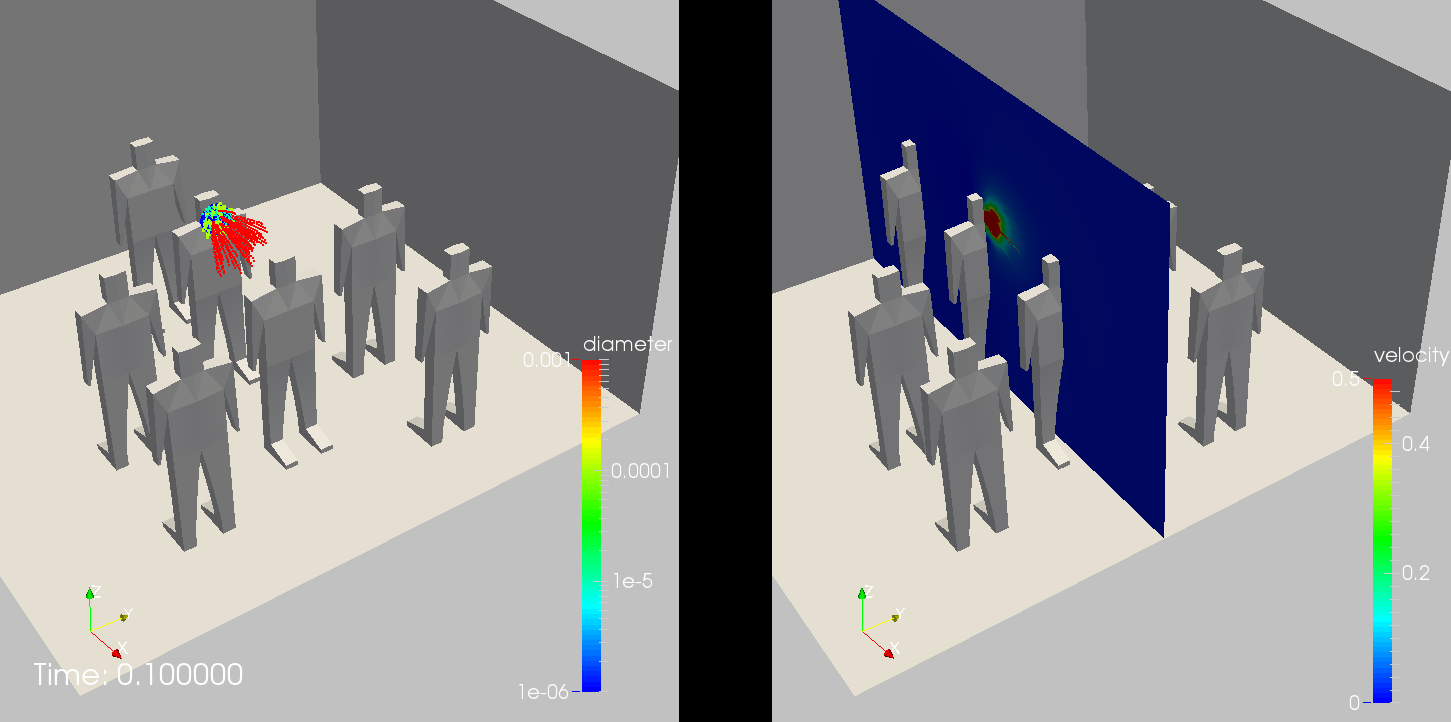}
	\caption{TSA Queue: Particle Distribution at $t=0.10~sec$}
\end{figure}

\begin{figure}[h!]
\centering
	\includegraphics[width=10cm]{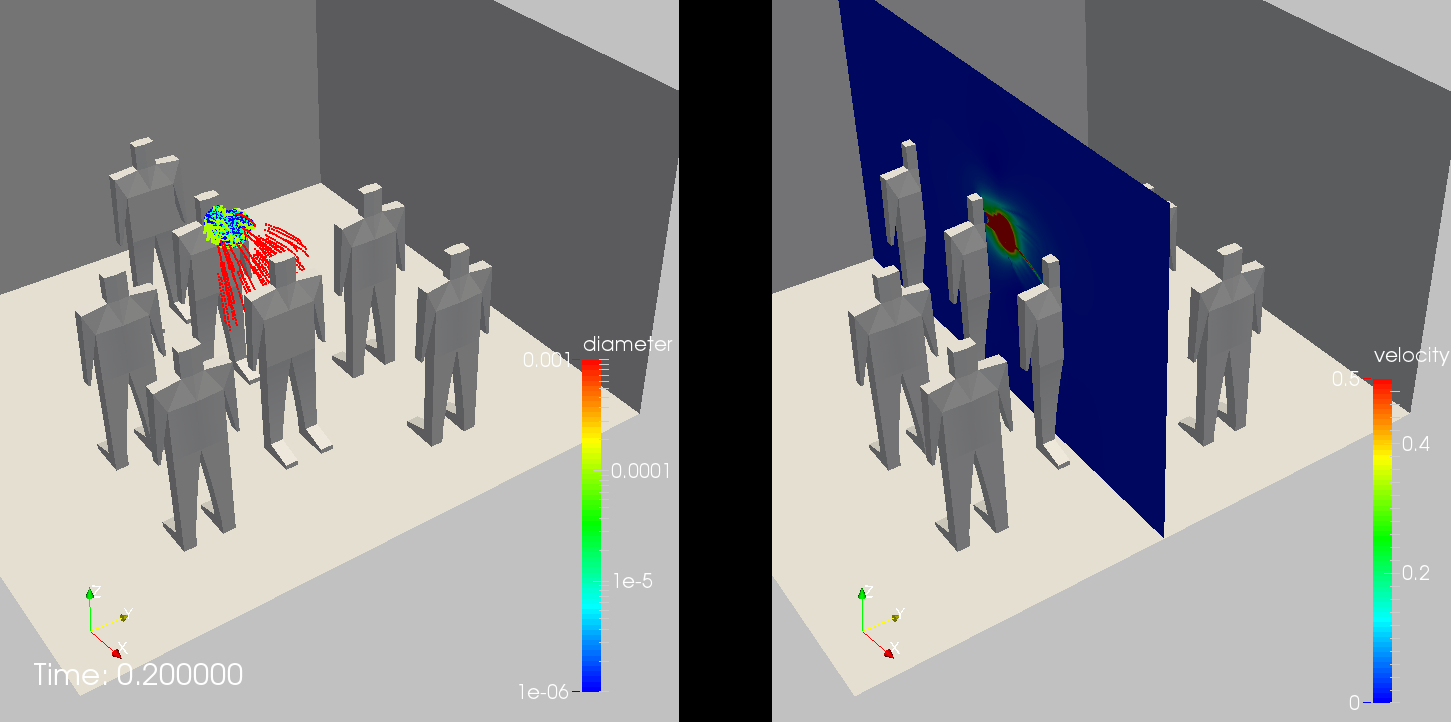}
	\caption{TSA Queue: Particle Distribution at $t=0.20~sec$}
\end{figure}

\begin{figure}[h!]
\centering
	\includegraphics[width=10cm]{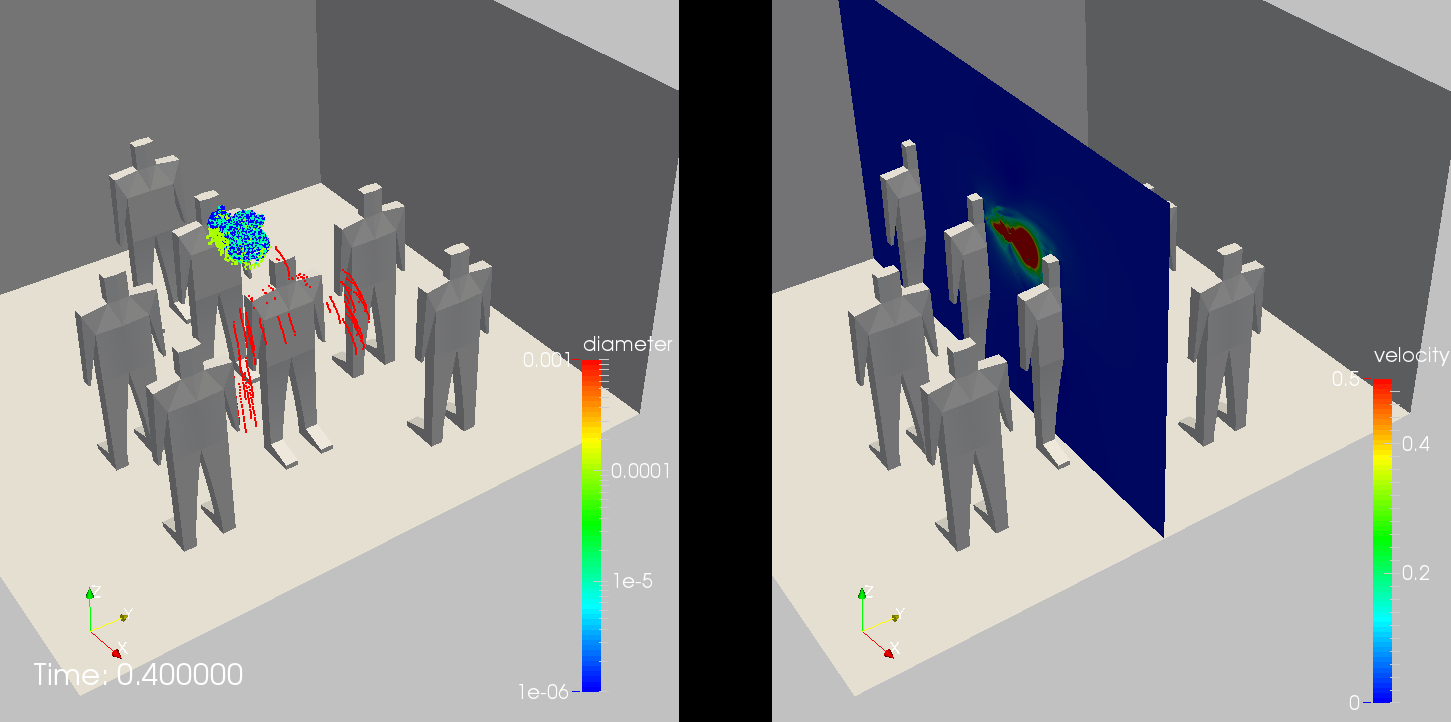}
	\caption{TSA Queue: Particle Distribution at $t=0.40~sec$}
\end{figure}

\begin{figure}[h!]
\centering
	\includegraphics[width=10cm]{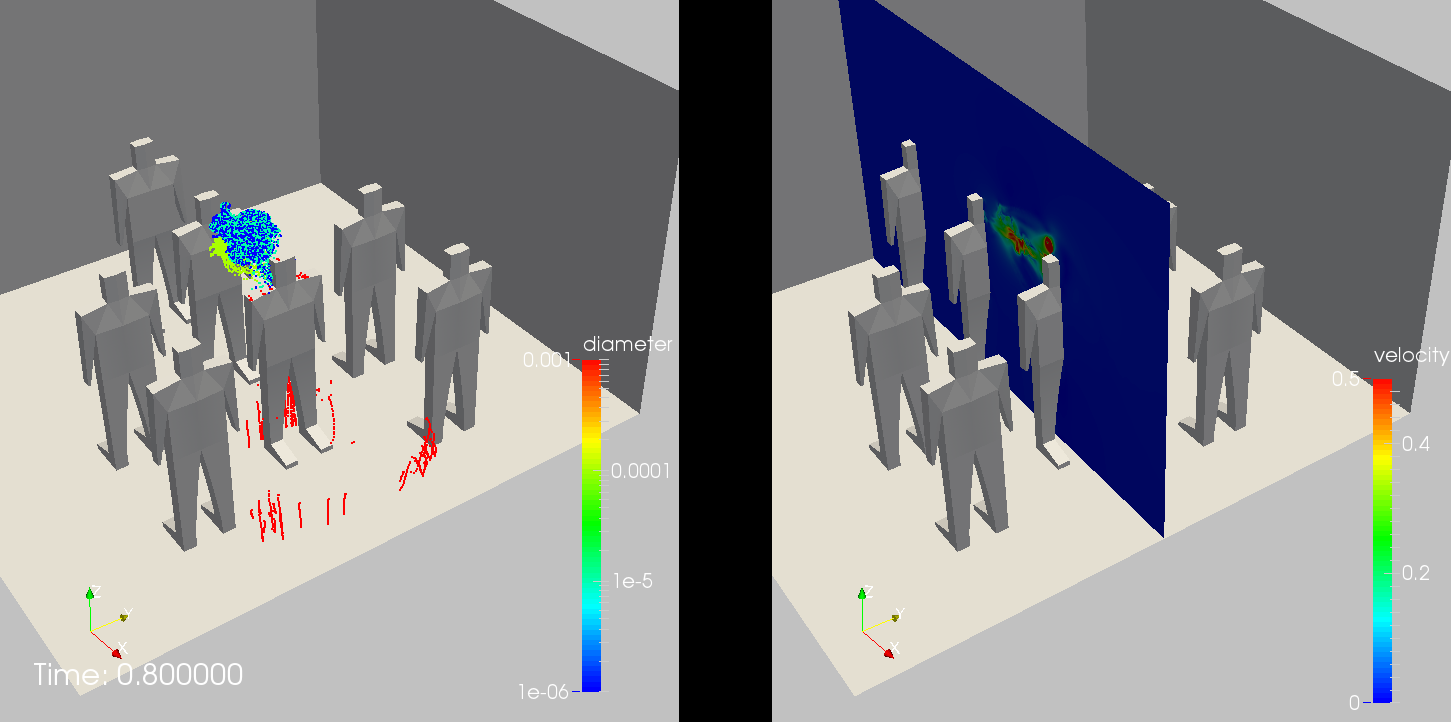}
	\caption{TSA Queue: Particle Distribution at $t=0.80~sec$}
\end{figure}

\begin{figure}[h!]
\centering
	\includegraphics[width=10cm]{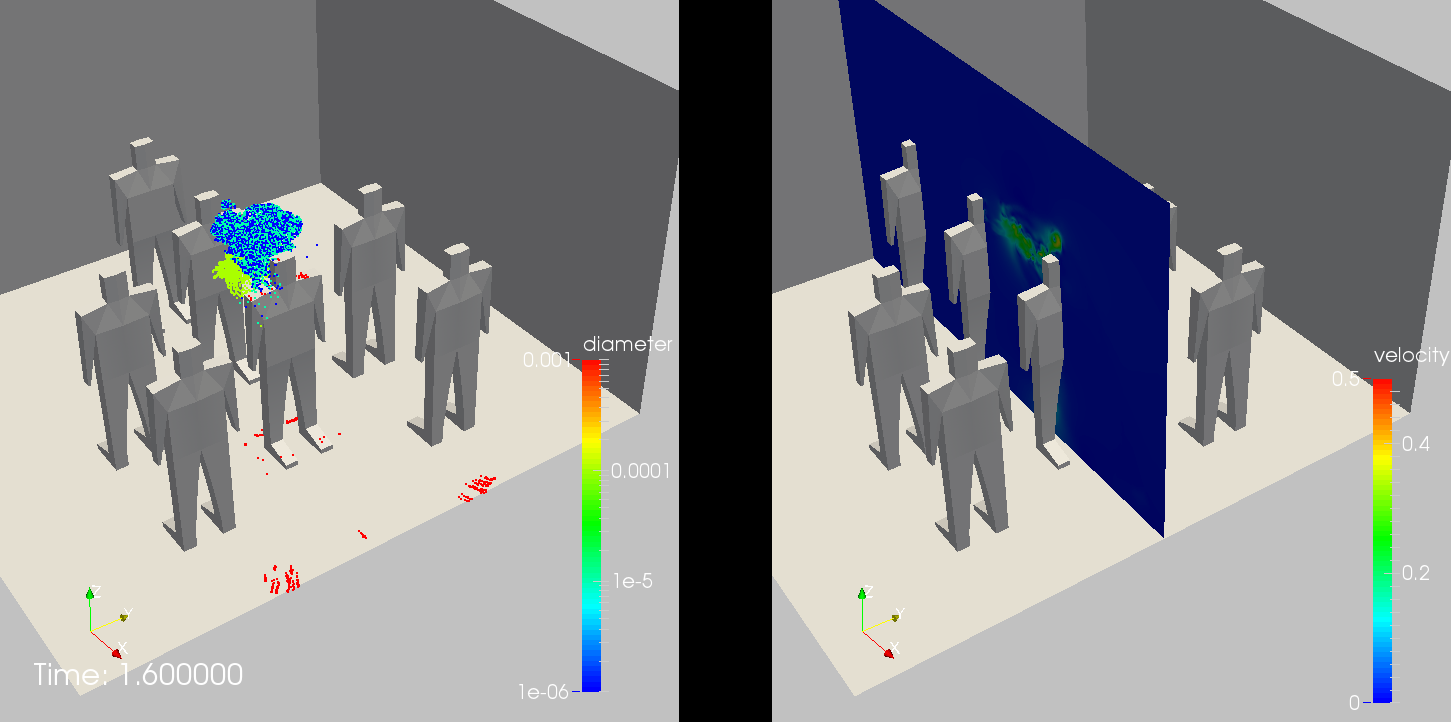}
	\caption{TSA Queue: Particle Distribution at $t=1.60~sec$}
\end{figure}

\subsection{Sneezing in a Generic Hospital Room}
This case considers a typical hospital room. Of interest here was the
dispersion of particles in the {\bf first minute} after coughing, in
particular the reach into neighbouring halls and the amount of `negative
pressure' needed to keep all contaminants in the room. 
Figure~11 shows the arrangement of the room, with patient and caregiver
clearly visible. This particular mesh had {\code 2.25Mels}.
The distribution of particles over time can be discerned from
Figures~12-22. As before, one can see that the large (red) particles follow 
a ballistic path.
This `ballistic phase' ends at about $t=1~sec$. The (green) particles of size
$d=0.1~mm$ are quickly stopped by the air, and then sink slowly towards
the patient. The even smaller
(cyan, blue) particles rise with the cloud of warmer air exhaled by the
sneezing individual, and disperse much further at later times, covering
almost the entire room. The velocity distribution in the
room may be inferred from Figure~23.

\begin{figure}[h!]
\centering
	\includegraphics[width=10cm]{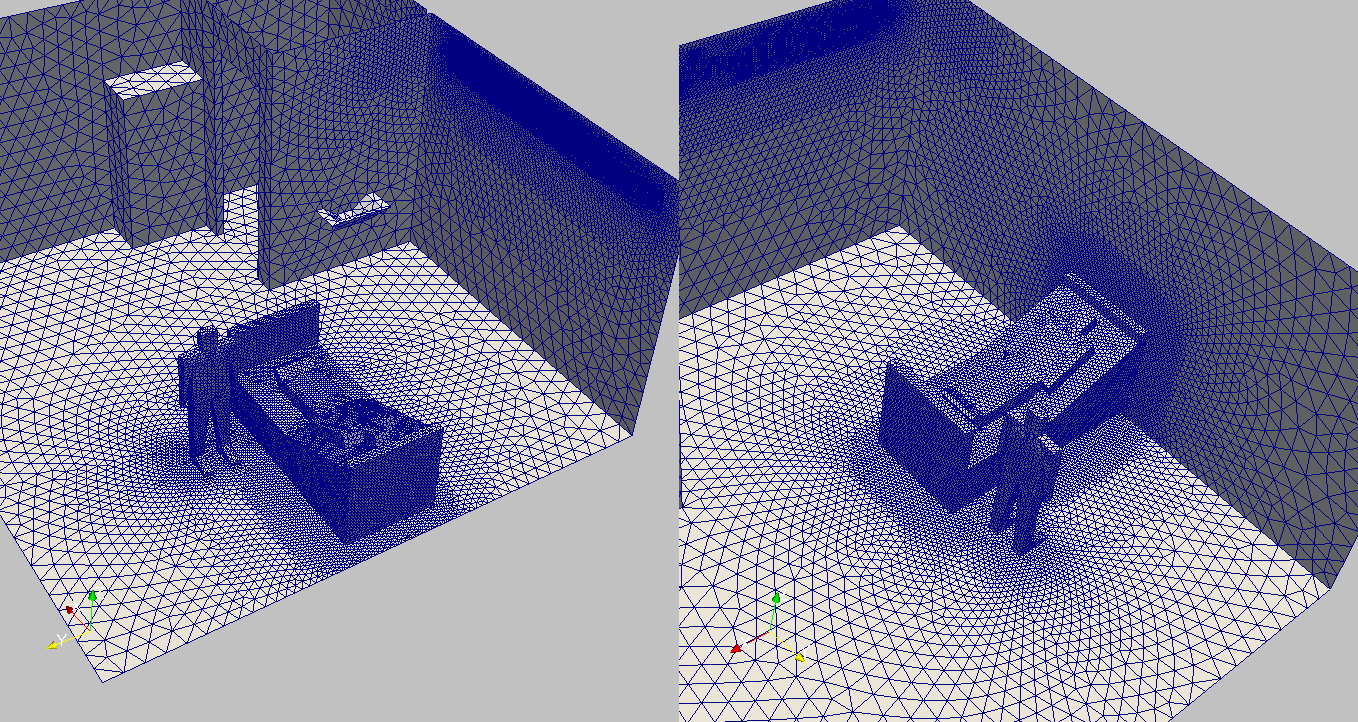}
	\caption{Hospital Room: Surface Mesh}
\end{figure}

\begin{figure}[h!]
\centering
	\includegraphics[width=10cm]{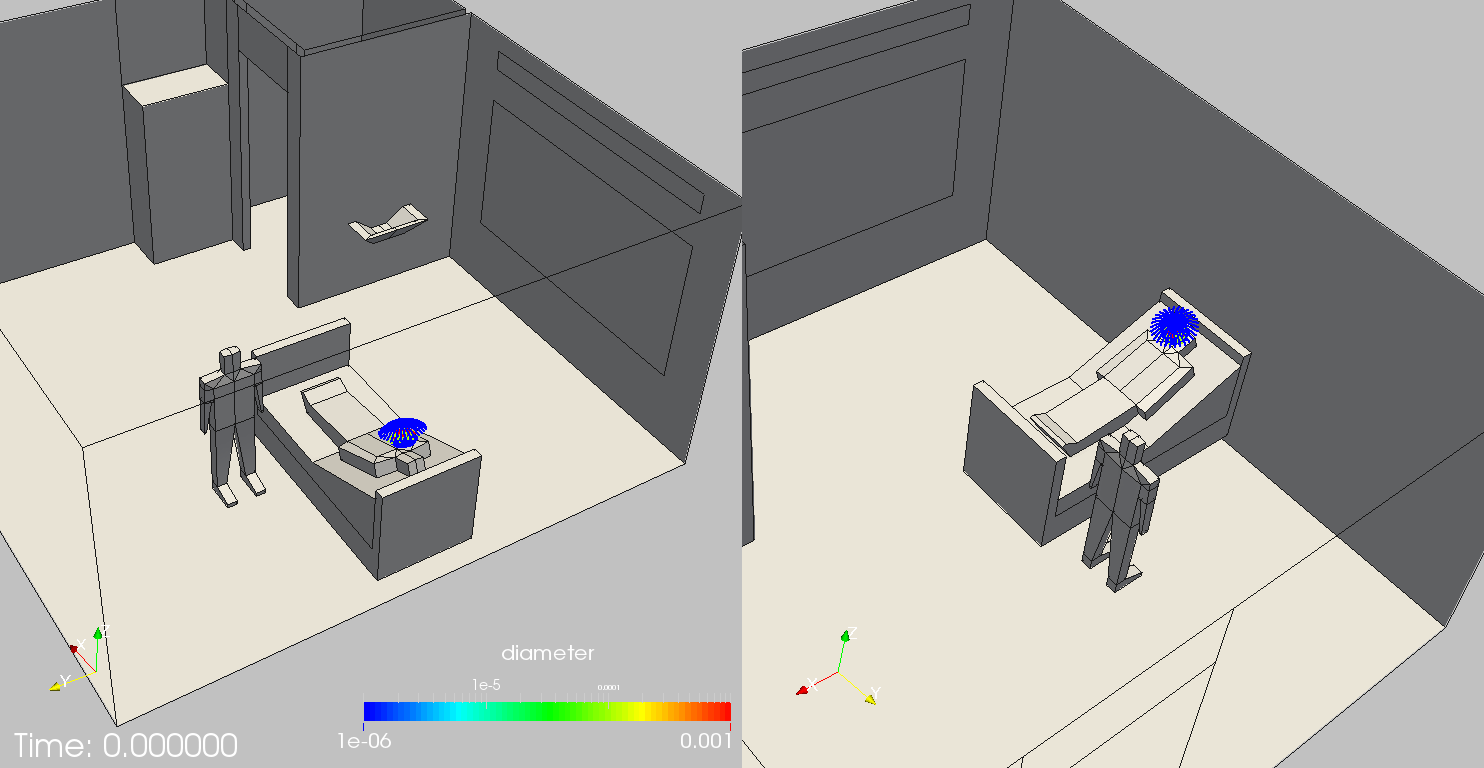}
	\caption{Hospital Room: Particle Distribution at t=0.0 sec}
\end{figure}

\begin{figure}[h!]
\centering
	\includegraphics[width=10cm]{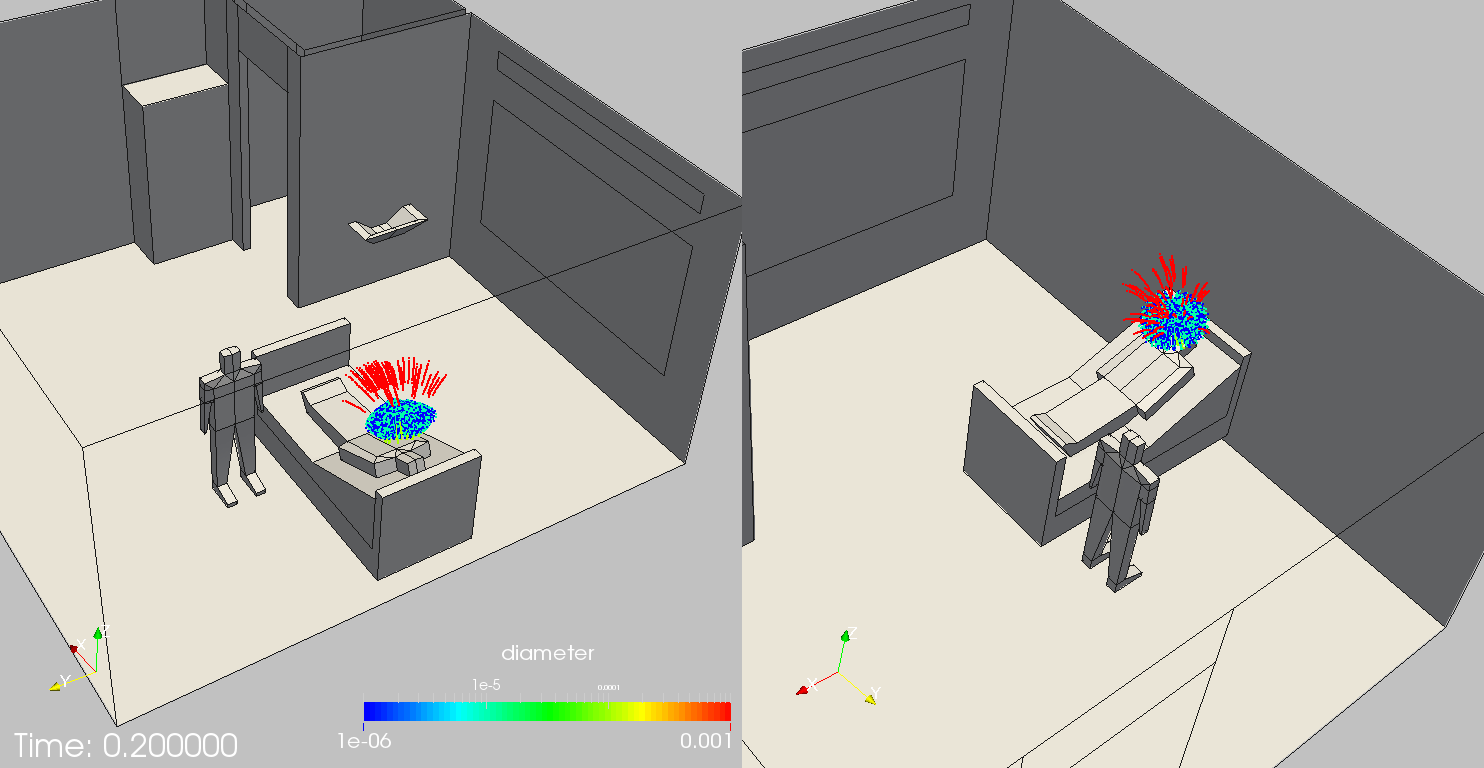}
	\caption{Hospital Room: Particle Distribution at t=0.2 sec}
\end{figure}

\begin{figure}[h!]
\centering
	\includegraphics[width=10cm]{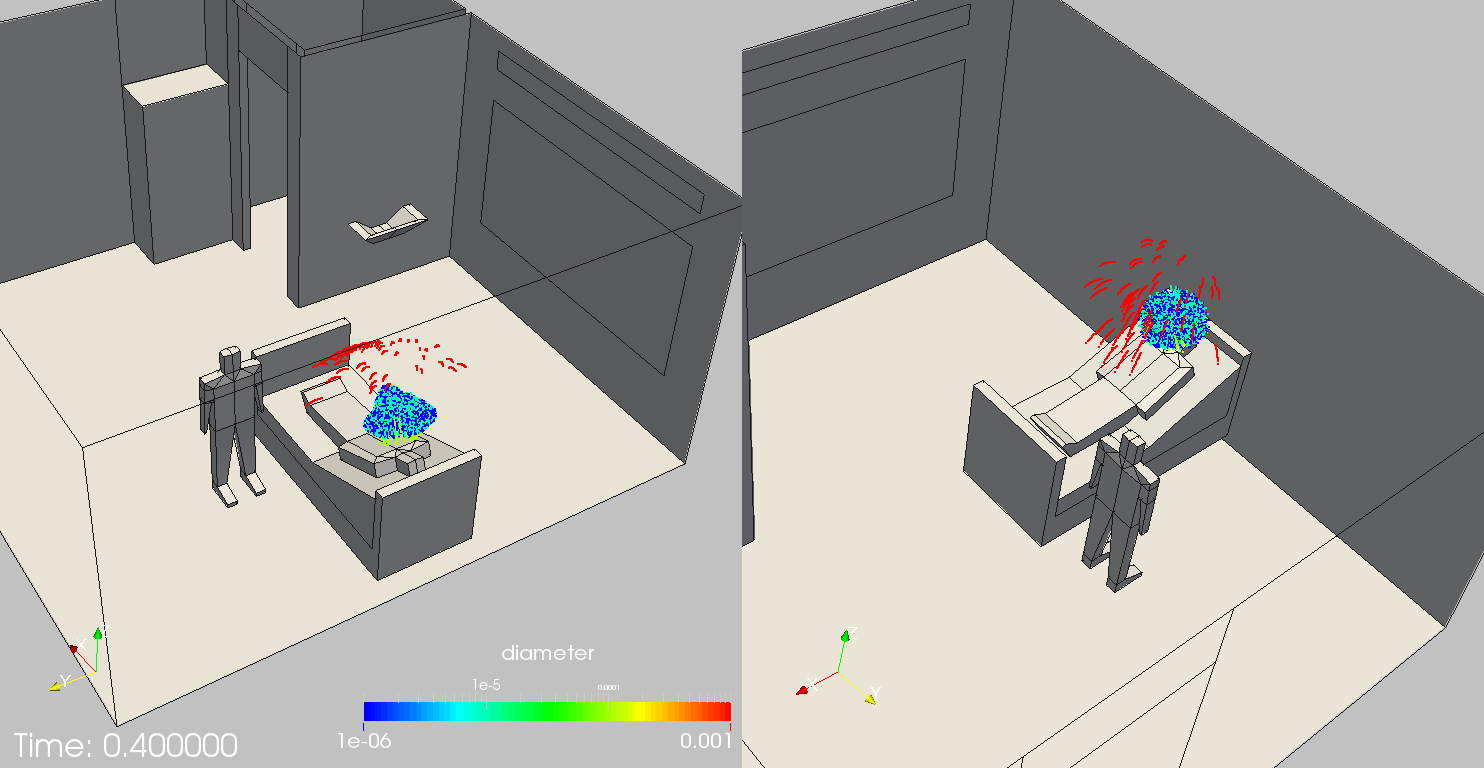}
	\caption{Hospital Room: Particle Distribution at t=0.4 sec}
\end{figure}

\begin{figure}[h!]
\centering
	\includegraphics[width=10cm]{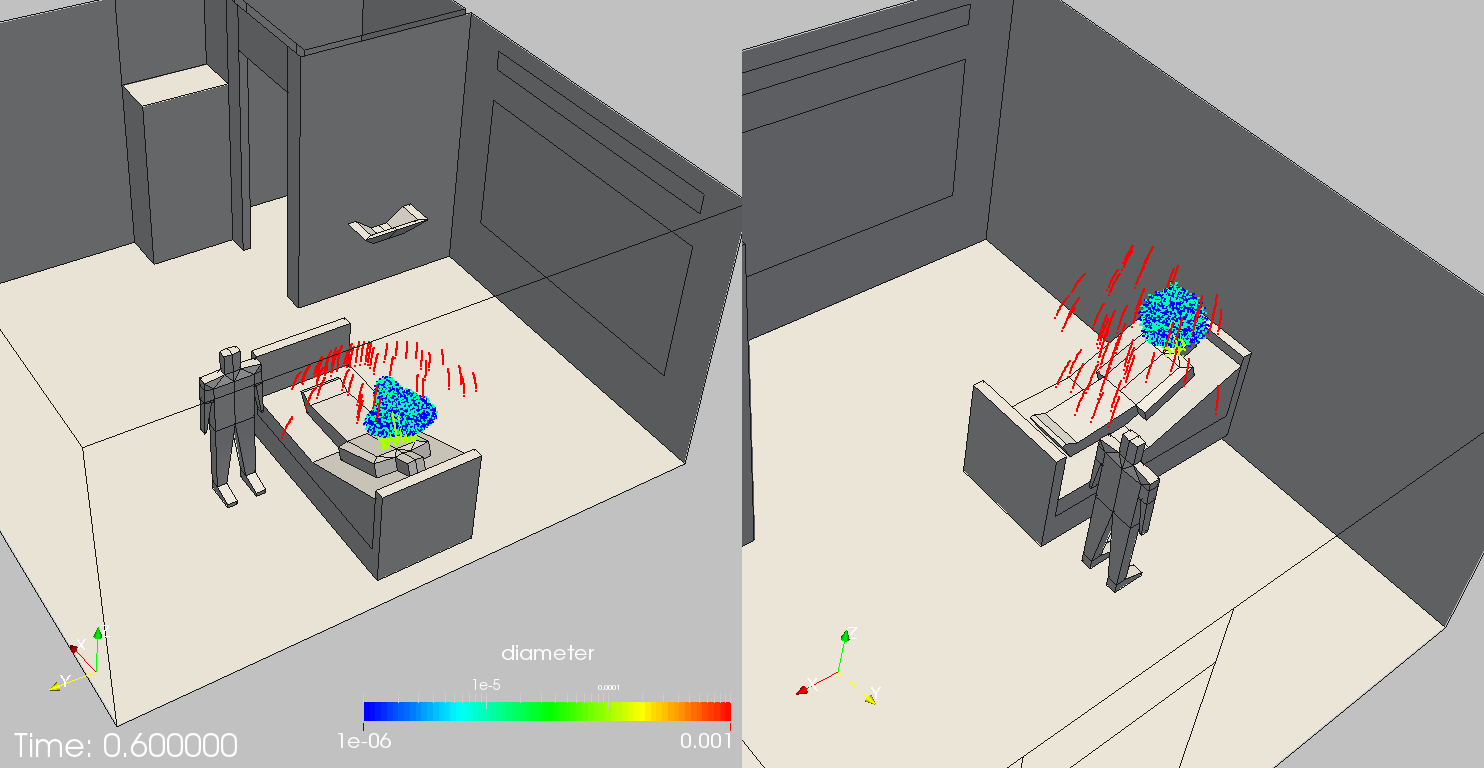}
	\caption{Hospital Room: Particle Distribution at t=0.6 sec}
\end{figure}

\begin{figure}[h!]
\centering
	\includegraphics[width=10cm]{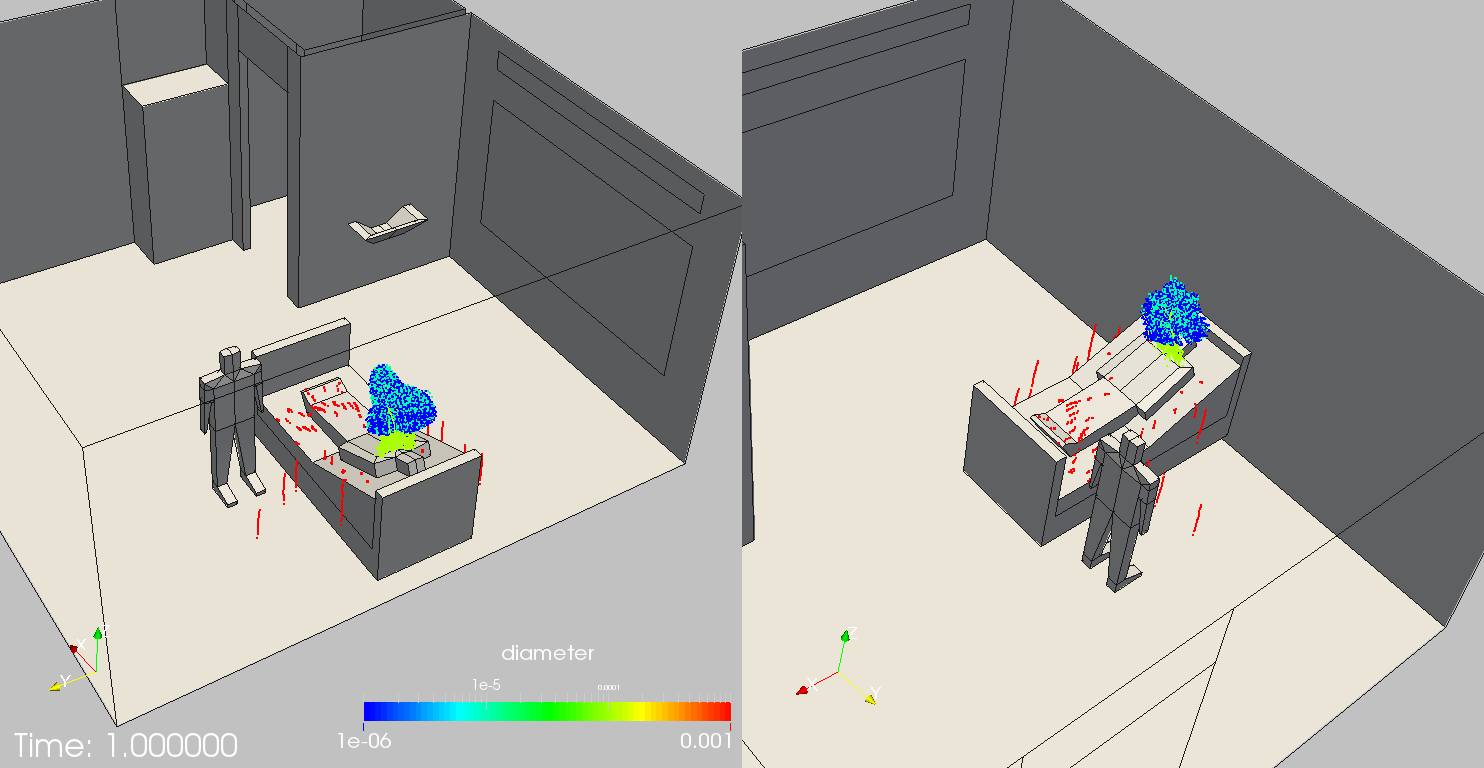}
	\caption{Hospital Room: Particle Distribution at t=1.0 sec}
\end{figure}

\begin{figure}[h!]
\centering
	\includegraphics[width=10cm]{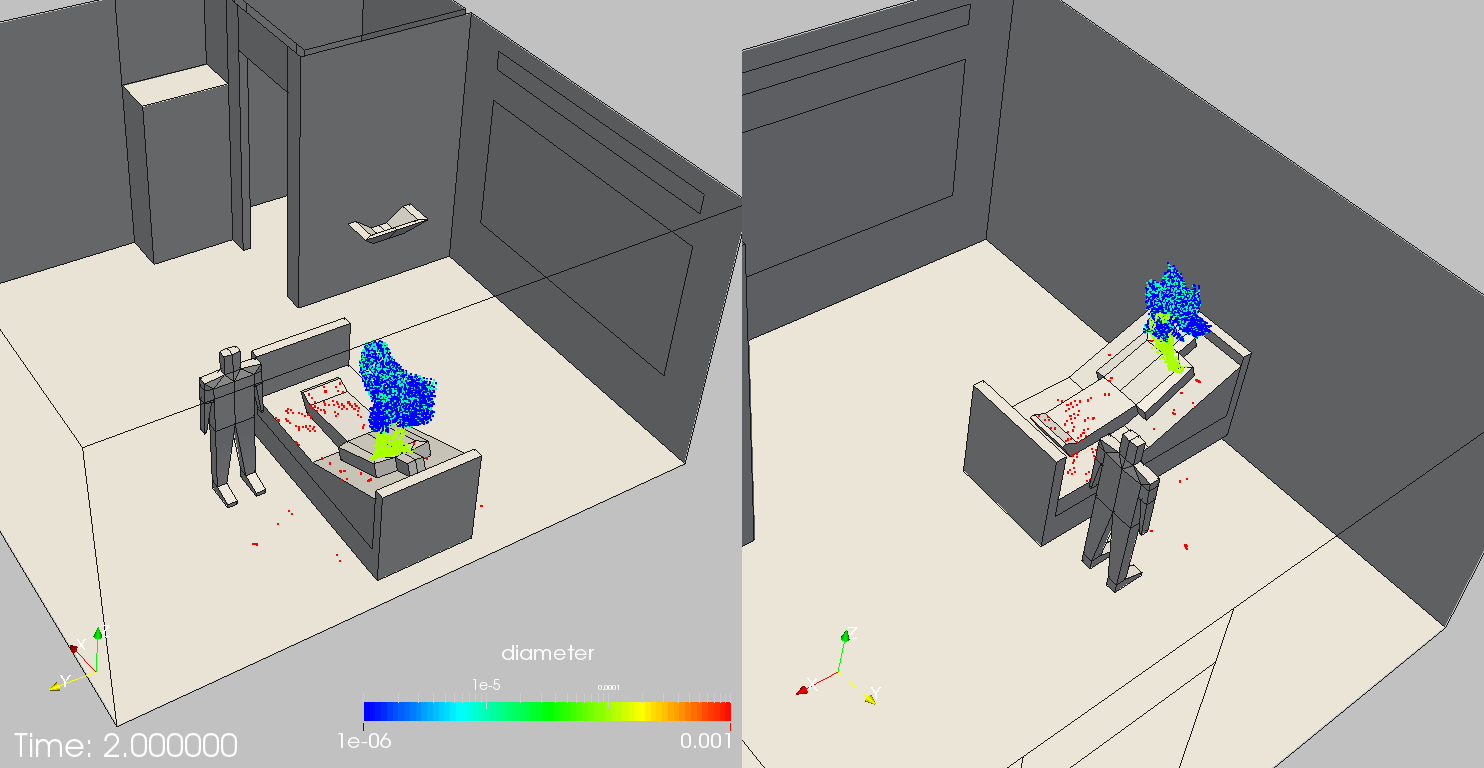}
	\caption{Hospital Room: Particle Distribution at t=2.0 sec}
\end{figure}

\begin{figure}[h!]
\centering
	\includegraphics[width=10cm]{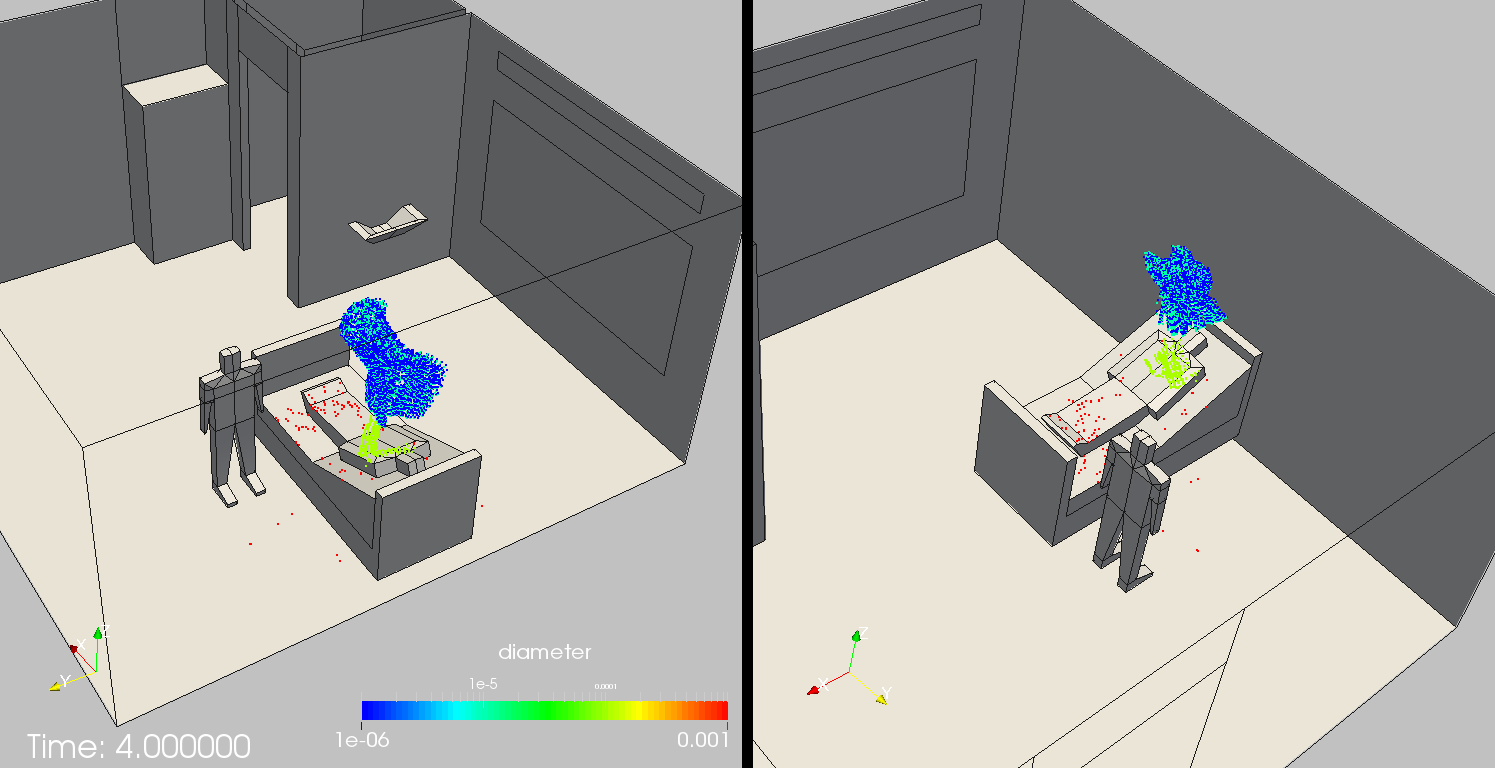}
	\caption{Hospital Room: Particle Distribution at t=4.0 sec}
\end{figure}

\begin{figure}[h!]
\centering
	\includegraphics[width=10cm]{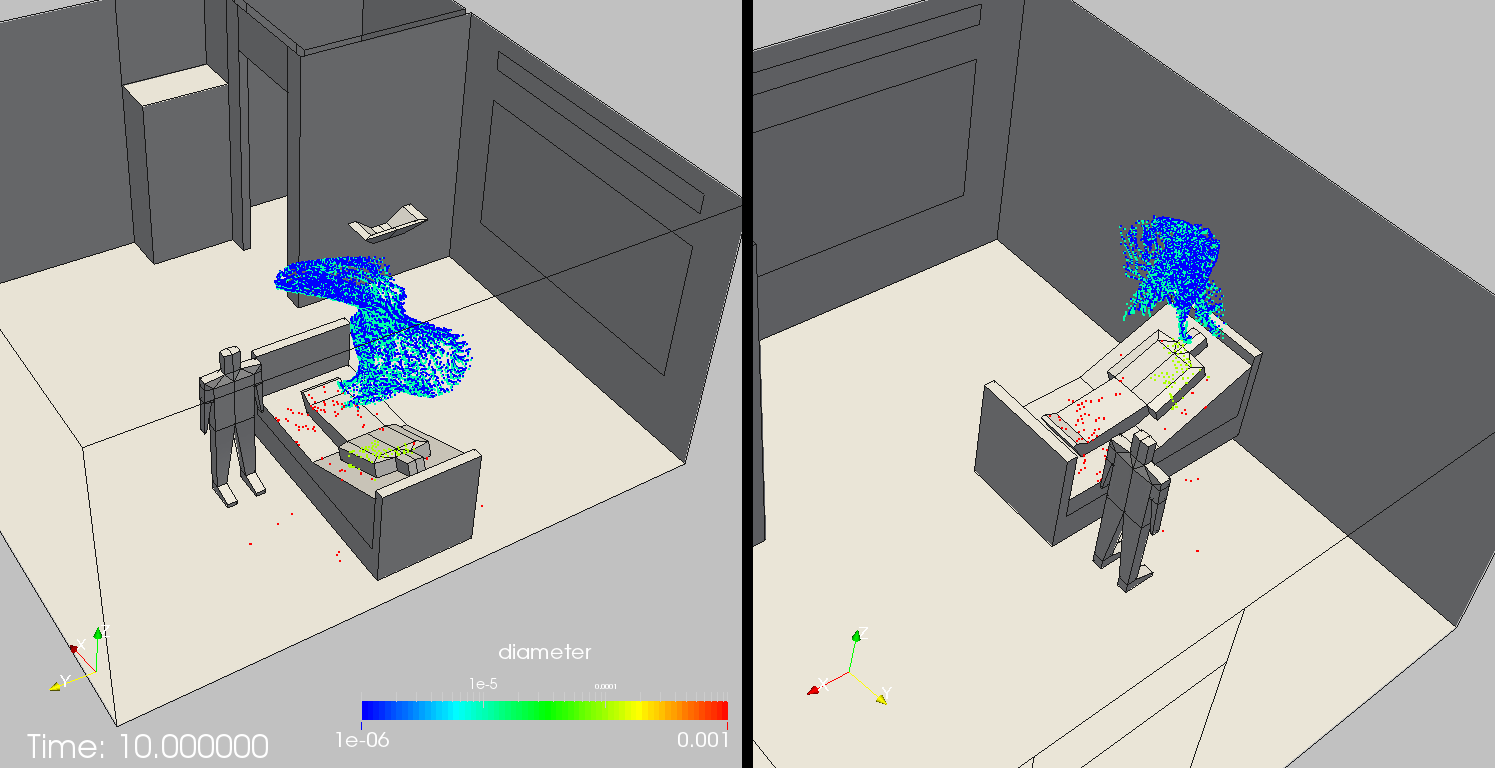}
	\caption{Hospital Room: Particle Distribution at t=10.0 sec}
\end{figure}

\begin{figure}[h!]
\centering
	\includegraphics[width=10cm]{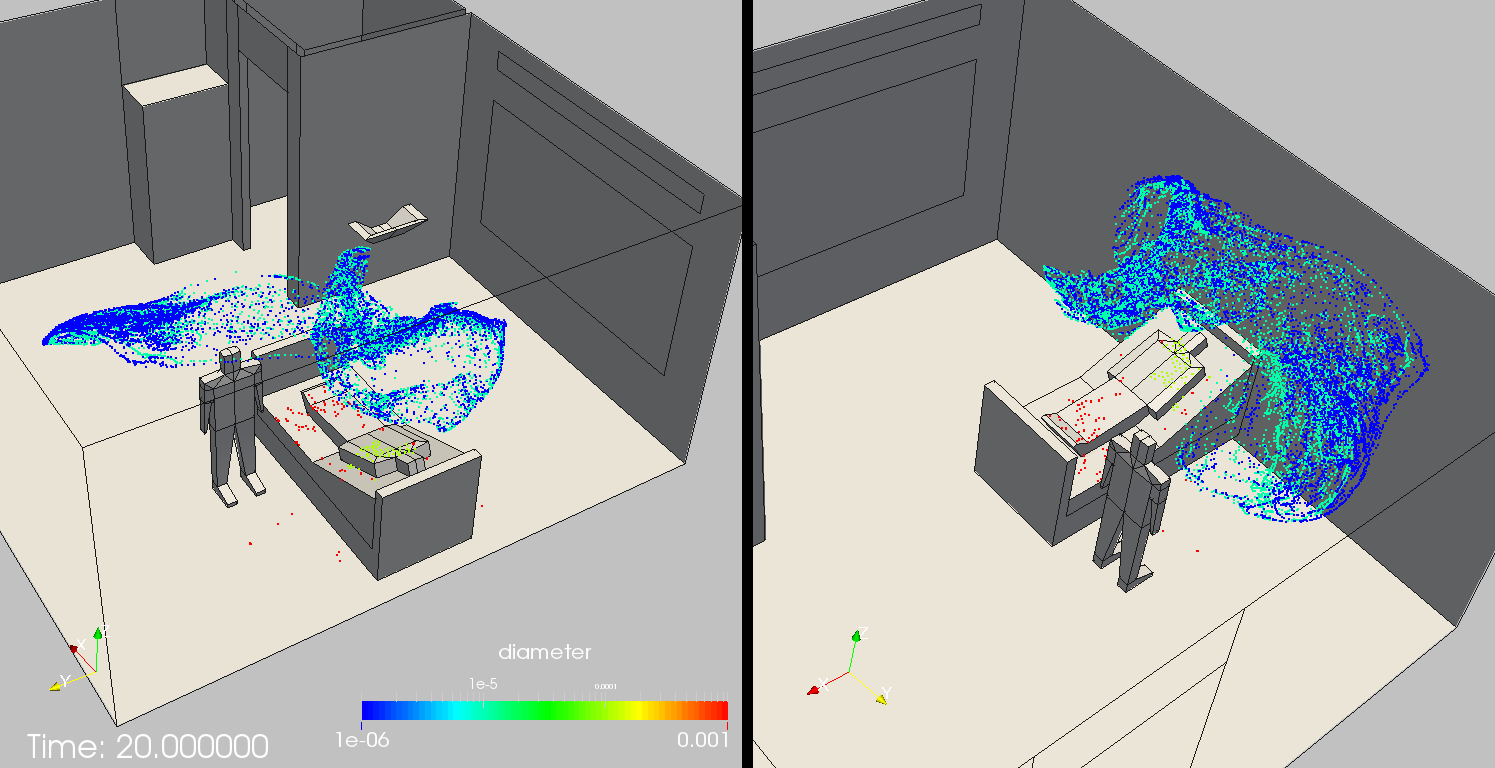}
	\caption{Hospital Room: Particle Distribution at t=20.0 sec}
\end{figure}

\begin{figure}[h!]
\centering
	\includegraphics[width=10cm]{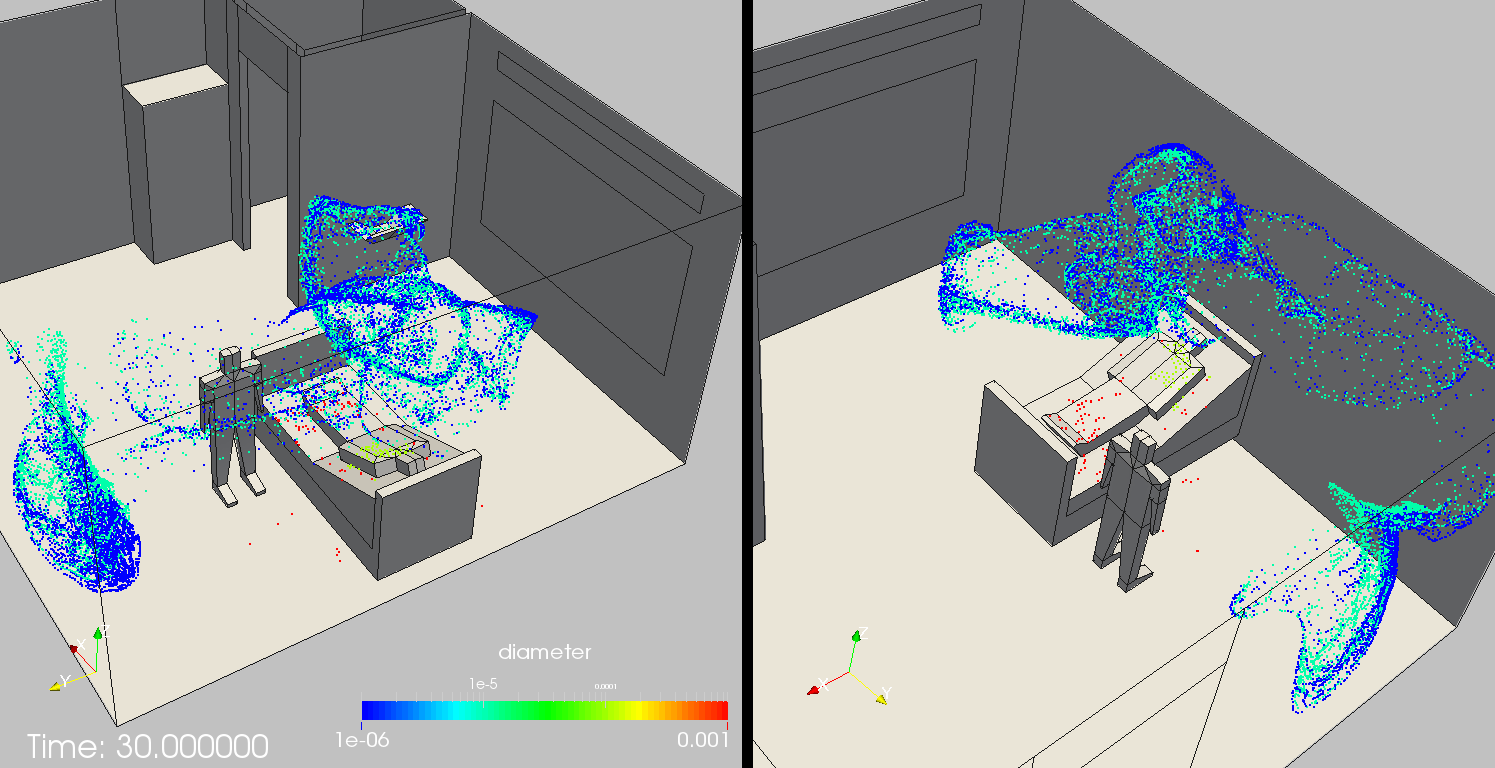}
	\caption{Hospital Room: Particle Distribution at t=30.0 sec}
\end{figure}

\begin{figure}[h!]
\centering
	\includegraphics[width=10cm]{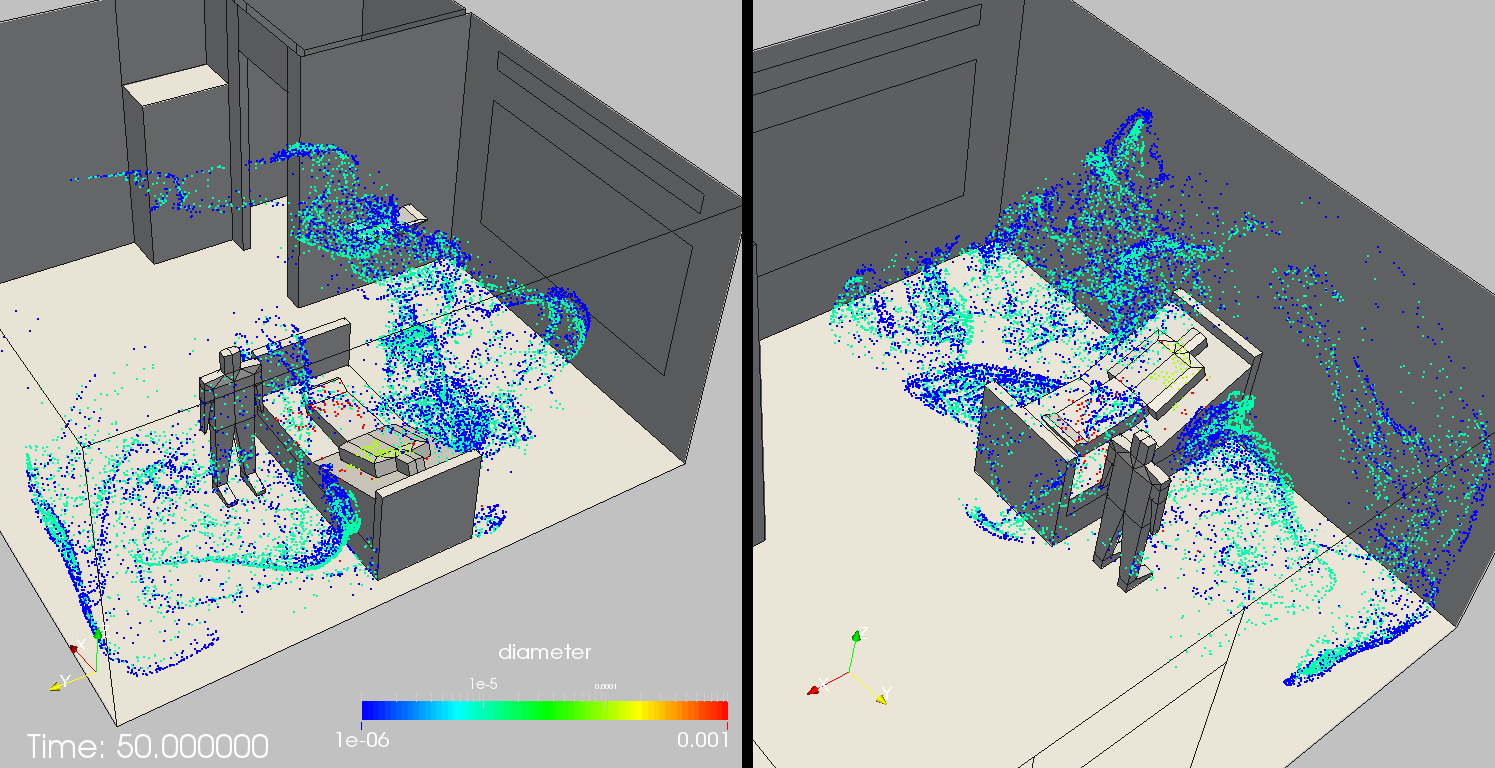}
	\caption{Hospital Room: Particle Distribution at t=50.0 sec}
\end{figure}

\begin{figure}[h!]
\centering
	\includegraphics[width=10cm]{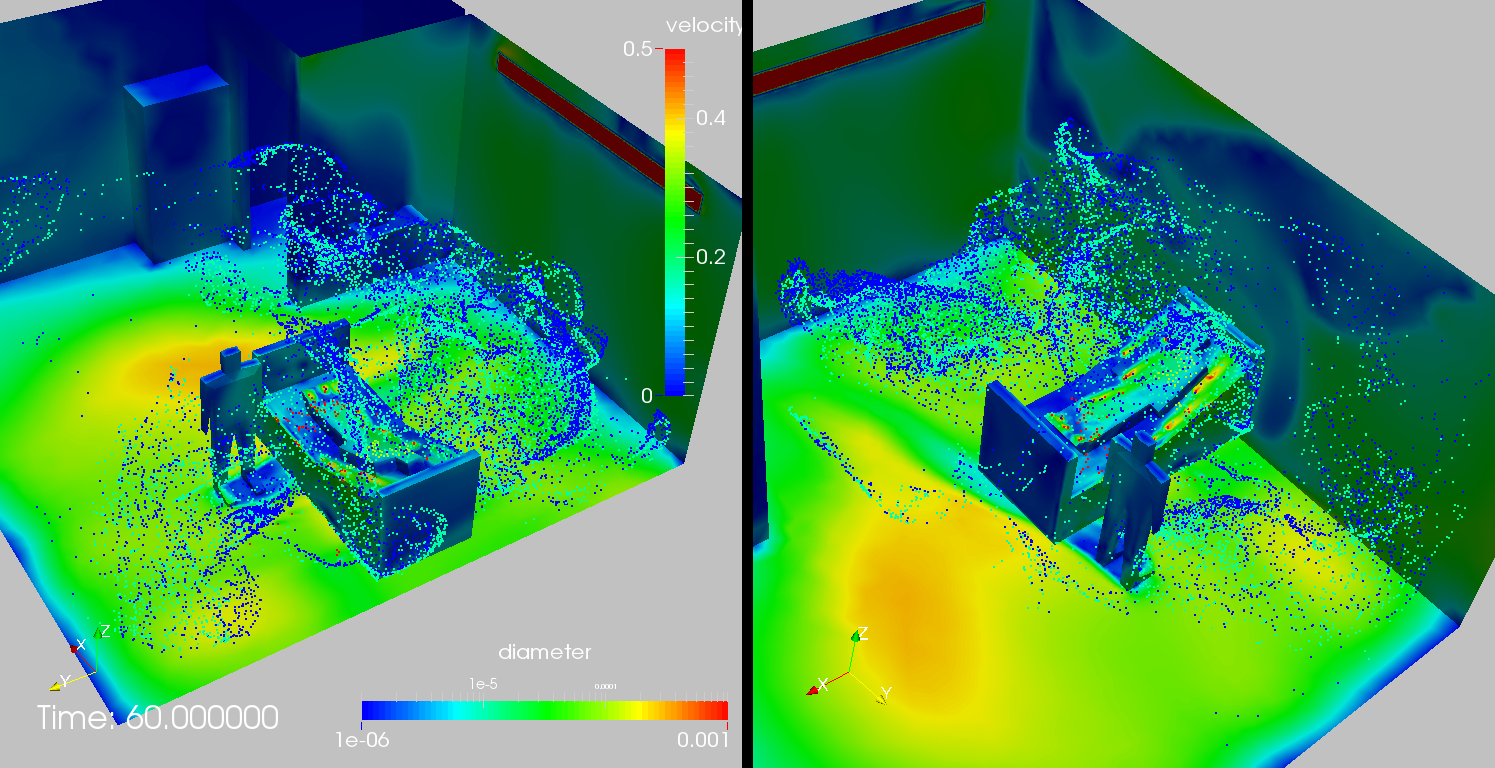}
	\caption{Hospital Room: Surface Velocities at t=60.0 sec}
\end{figure}

\section{Reopening After the Crisis}
A lingering question facing all levels of society is how and when to
reopen facilities where people congregate in close proximity. One key
technology that would allow opening is testing and sensing. We consider
sensing in the sequel. Several vendors have announced measuring devices
for Covid-19 in the next half year. Given that these sensors are expensive,
and that a hospital or university many need hundreds of these, the question
becomes how best to deploy them. In other words: given an arbitrary number
of contamination or infection scenarios, which is the minimum number of
sensors needed to detect them, and where should they be placed~?
A partial answer to this non-trivial question was given in 
\cite{Loh05,Zha07}. 
If we assume a given number of sensors, every contaminant/infection 
scenario (location and amount of release, flow conditions, etc.) will 
lead to a sensor input.
The data recorded from all the possible release scenarios at all possible
sensor locations allows the identification of the best or optimal sensor
locations. Clearly, if only one sensor is to be placed, it should be at
the location that recorded the highest number of releases. This argument
can be used recursively by removing from further consideration all
releases already recorded by sensors previously placed. 
The procedure is repeated recursively until no undetected release cases 
are left, or the available sensors have been exhausted.  \\
See \cite{Bur15,Hin17} for an in-depth analysis of robust sensor 
placement under uncertainty.

\subsection{Hospital Room}
This case considers the same hospital room as shown before. The 
boundary conditions determining the flow are assumed as steady,
with air entering the room through vents~1-3 and exiting the room
through the bathroom exhaust or the door. Figures~24-26 show the
outlay of the room, average velocities and the `age of air' after
5~minutes. Note the high values for the age of air in the corners
and the back of the room. This particular mesh had {\code 2.2Mels}.
Four contaminant release scenarios were
considered: cases~1-3 assumed contaminant coming in through each
of the vents (separately) during the first minute, while
case~4 assumed virus production from the patient for a period
of 10~seconds. The case was run for 5~minutes of real time, and
the contaminant concentration was measured on all walls/ceilings.
The maximum concentrations measured have been summarized in
Figure~27. Note the different areas covered depending on the release
scenario. It was assumed that sensors should only be allowed above a
certain height, and should be located on a wall or the ceiling.
Table~4 summarizes the points that measured data above a set
threshold. As one can see, none of the possible sensor locations
is able to measure/detect all 4~cases, and many possible sensor
locations do not detect even a single case. There are many possible
pairs of sensors that can detect all 4~cases. The pair selected
is the one that achieves the highest relative measurement values,
and is shown in Figure~28. Note that this makes good sense: one sensor
close the HVAC exits, and one close to the patient.

\begin{figure}[h!]
\centering
	\includegraphics[width=10cm]{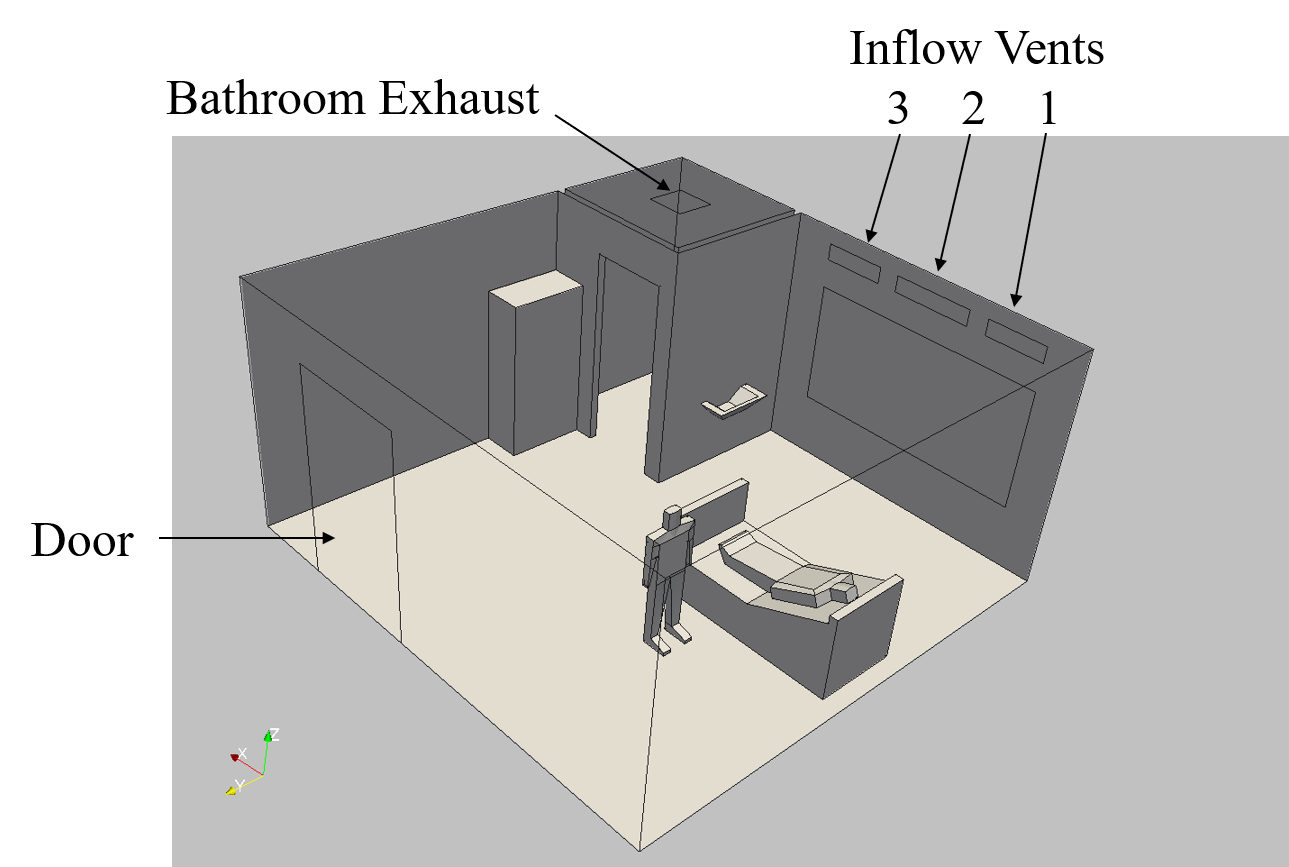}
	\caption{Hospital Room: Outlay of Room and Boundary
Conditions}
\end{figure}

\begin{figure}[h!]
\centering
	\includegraphics[width=10cm]{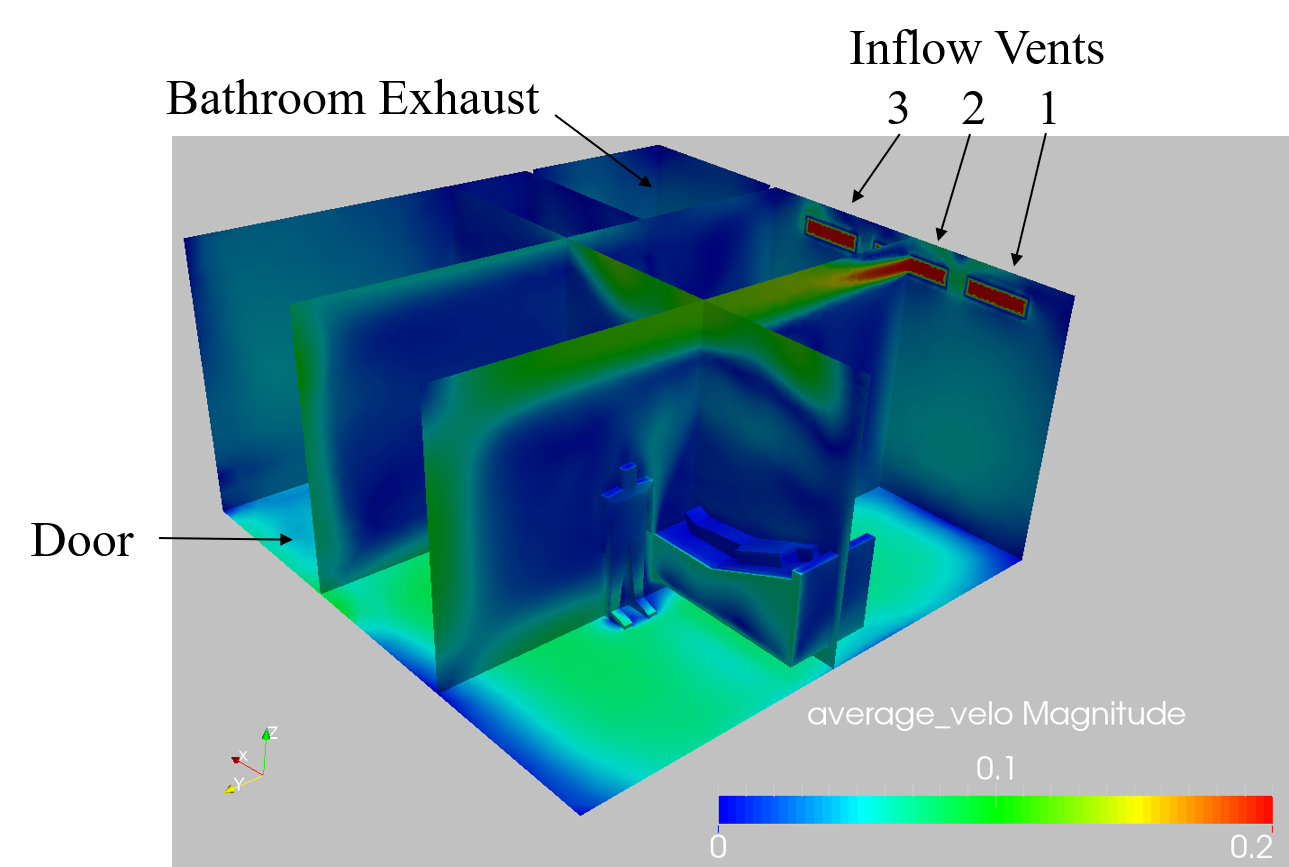}
	\caption{Hospital Room: Average Velocities (5~mins)}
\end{figure}

\begin{figure}[h!]
\centering
	\includegraphics[width=10cm]{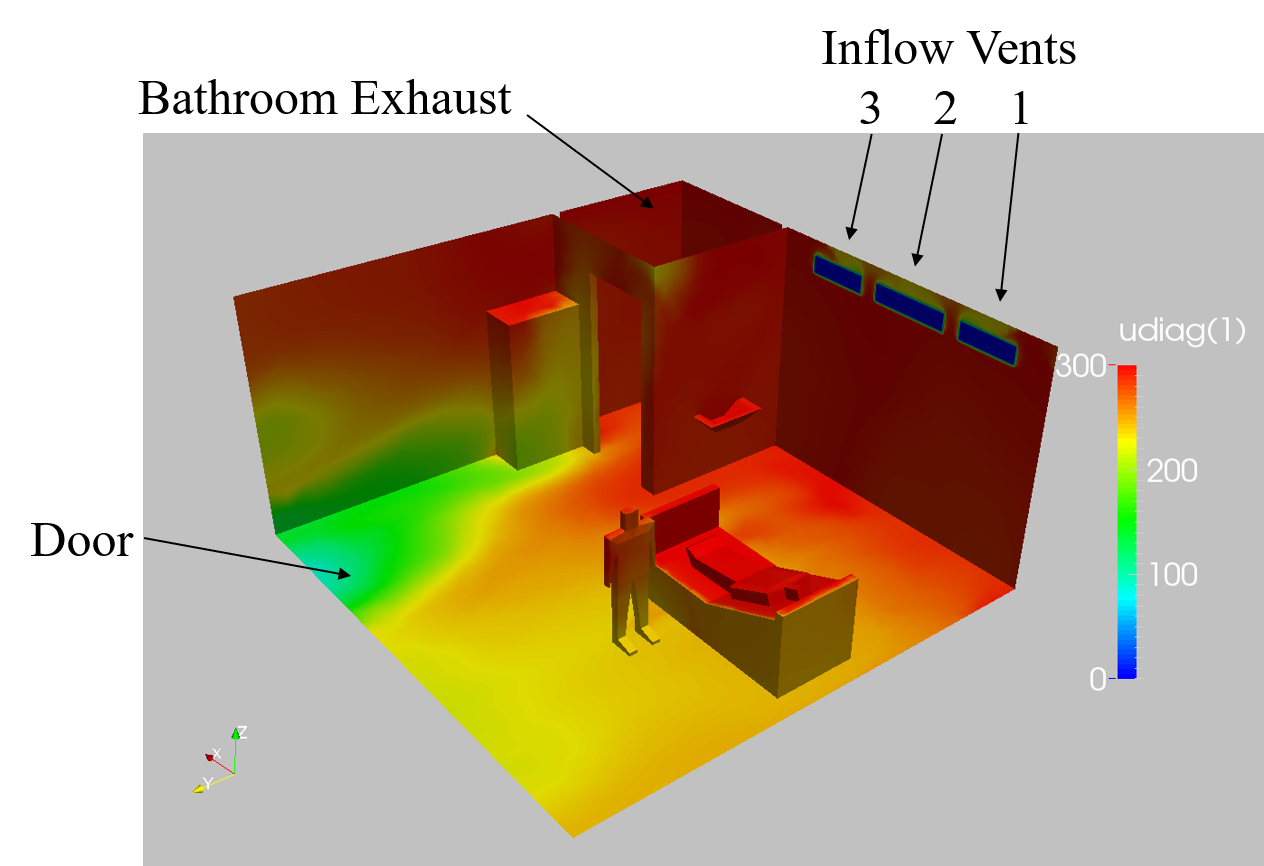}
	\caption{Hospital Room: Age of Air (5~mins)}
\end{figure}

\begin{figure}[h!]
\centering
	\includegraphics[width=10cm]{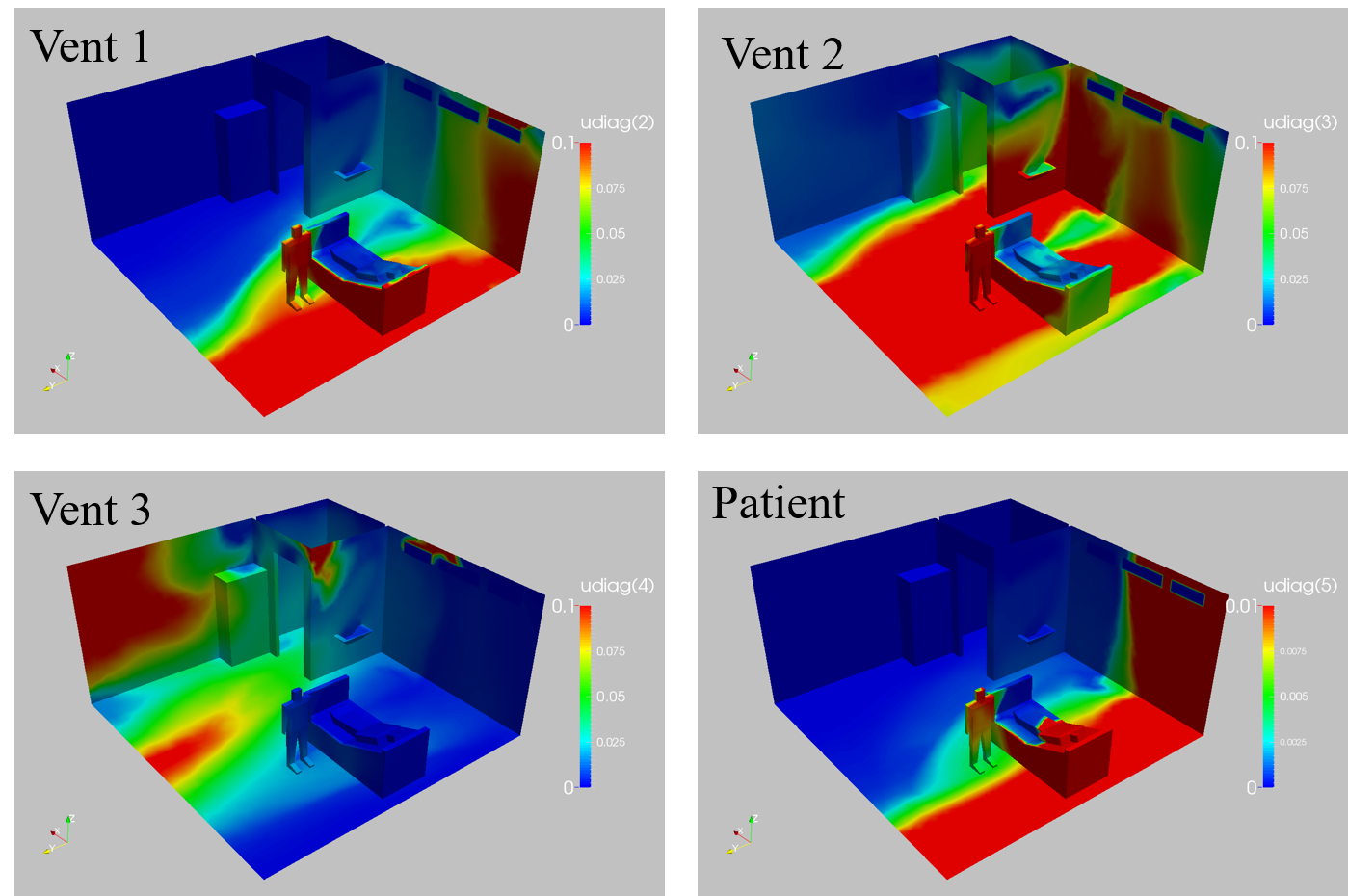}
	\caption{Hospital Room: Maximum Contaminant Concentration
Over 5 Minutes}
\end{figure}

\begin{table}[htbp]
\begin{center}
\caption{Data Measurement Summary}
\label{tab:meas_data}
\begin{tabular}{c|c}
\hline
Cases Measured & Number \\
\hline
0 & 4308 \\
1 & 3377 \\
2 & 1010 \\
3 &    0 \\
4 &    0 \\
\hline
\end{tabular}
\end{center}
\end{table}

\begin{figure}[h!]
\centering
	\includegraphics[width=10cm]{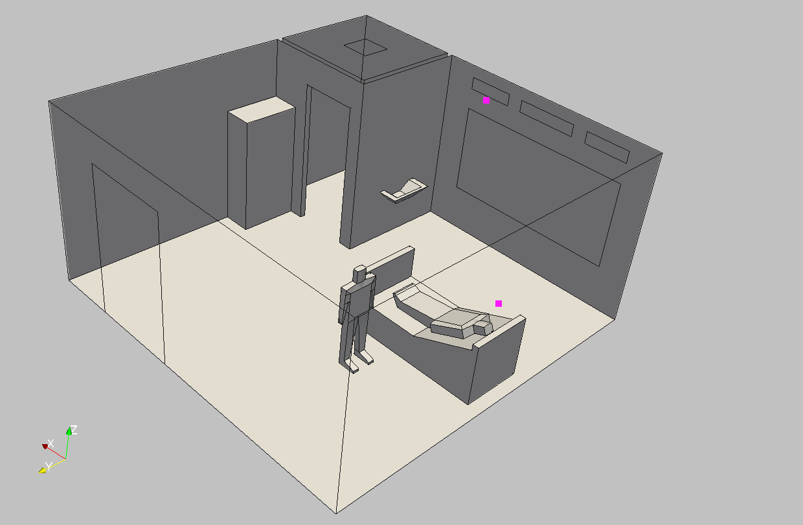}
	\caption{Hospital Room: Optimal Sensor Locations}
\end{figure}

\section{Conclusions and Outlook}
\label{sec:conclusions}
The present paper has summarized some of the mechanical characteristics
of virus contaminants and the transmission via droplets and aerosols.
The ordinary and partial differential equations describing the
physics of these processes with high fidelity
were given, as well as appropriate numerical schemes to solve them.
Several examples taken from recent evaluations of the built environment 
were given, as well as the optimal placement of sensors.

Current efforts are directed at increasing the realism of the
physical processes modeled (e.g. by adding the effect of moving
pedestrians \cite{Loh16}), streamlining the simulation toolbox
and workflow, and fielding these tools so that the post-pandemic 
opening can occur as smoothly as possible.

\section*{Acknowledgements}

The authors would like to thank Carlos N. Rautenberg for various discussions on 
Section~7.1.


\bibliographystyle{aiaa}

\begin{thebibliography}{7}

\bibitem{App08} S. Appanaboyina, F. Mut, R. L\"ohner, C. M. Putman and
J. R. Cebral - Computational Fluid Dynamics of Stented Intracranial
Aneurysms Using Adaptive Embedded Unstructured Grids;
{\sl Int.\ J.\ Num.\ Meth.\ Fluids } 57, 5, 475-493 (2008).

\bibitem{Asa19} S. Asadi, A.S. Wexler, C.D. Cappa, S. Barreda, N.M. Bouvier
and W. Ristenpart - Aerosol Emission and Superemission During Human
Speech Increase with Voice Loudness;
{\sl Nature Scientific Reports }9, (1):2348 (2019).
www.nature.com/scientificreports/
https://doi.org/10.1038/s41598-019-38808-z

\bibitem{Asa20a} S. Asadi, A.S. Wexler, C.D. Cappa, S. Barreda, N.M. Bouvier
and W. Ristenpart - Effect of
Voicing and Articulation Manner on Aerosol Particle
Emission During Human Speech {\sl PLoS ONE} 15(1):e0227699 (2020).
https://doi.org/10.1371/journal.pone.0227699

\bibitem{Asa20b} S. Asadi, N.M. Bouvier, A.S. Wexler and W. Ristenpart -
The Coronavirus Pandemic and Aerosols: Does
COVID-19 Transmit via Expiratory Particles ?
{\sl Aerosol Science and Technology } (2020).
https://doi.org/10.1080/02786826.2020.1749229

\bibitem{Aub08} R. Aubry, F. Mut, R. L\"ohner and J. R. Cebral -
Deflated Preconditioned Conjugate Gradient Solvers for the
Pressure-Poisson Equation; {\sl J.\ Comp.\ Phys.\ }227, 24,
10196-10208 (2008).

\bibitem{Bau91} J.D. Baum and R. L\"ohner - Numerical Simulation of Shock
Interaction with a Modern Main Battlefield Tank; {\sl AIAA}-91-1666 (1991).

\bibitem{Bau93} J.D. Baum and R. L\"ohner - Numerical Simulation of Pilot/Seat
Ejection from an F-16; {\sl AIAA}-93-0783 (1993).

\bibitem{Bau93a} J.D. Baum. H. Luo and R. L\"ohner - Numerical Simulation of a Blast
Inside a Boeing 747; {\sl AIAA}-93-3091 (1993).

\bibitem{Bau93b} J.D. Baum, H. Luo and R. L\"ohner - Numerical Simulation
of a Blast Withing a Multi-Room Shelter; pp. 451-463 in
{\sl Proc. MABS-13 Conf.} The Hague, Netherlands, September (1993).

\bibitem{Bau94} J.D. Baum, H. Luo and R. L\"ohner - A New ALE Adaptive
Unstructured Methodology for the Simulation of Moving Bodies;
{\sl AIAA}-94-0414 (1994).

\bibitem{Bau95} J.D. Baum, H. Luo and R. L\"ohner - Numerical Simulation of
Blast in the World Trade Center; {\sl AIAA}-95-0085 (1995).

\bibitem{Bau95a} J.D. Baum, H. Luo and R. L\"ohner - Validation of a New ALE,
Adaptive Unstructured Moving Body Methodology for Multi-Store Ejection
Simulations; {\sl AIAA}-95-1792 (1995).

\bibitem{Bau96} J.D. Baum, H. Luo, R. L\"ohner, C. Yang, D.
Pelessone and C. Charman - Coupled Fluid/Structure
Modeling of Shock Interaction with a Truck; {\sl AIAA}-96-0795 (1996).

\bibitem{Bau97} J.D. Baum, H. Luo, R. L\"ohner, E. Goldberg and
A. Feldhun - Application of Unstructured Adaptive
Moving Body Methodology to the Simulation of Fuel Tank Separation from
an F-16 c/d Fighter; {\sl AIAA}-97-0166 (1997).

\bibitem{Bau97a} J.D. Baum, R. L\"ohner, T.J. Marquette and H. Luo -
Numerical Simulation of Aircraft Canopy Trajectory; {\sl AIAA}-97-1885 (1997).

\bibitem{Bau99} J.D. Baum, H. Luo, E. Mestreau, R. L\"ohner,
D. Pelessone and C. Charman - Coupled CFD/CSD
Methodology for Modeling Weapon Detonation and Fragmentation;
AIAA-99-0794 (1999).

\bibitem{Bau06} J.D. Baum, E. Mestreau, H. Luo, R. L\"ohner,
D. Pelessone, M.E. Giltrud and J.K. Gran - Modeling of Near-Field Blast
Wave Evolution; {\sl AIAA}-06-0191 (2006).

\bibitem{Bal10} K. Balakrishnan and S. Menon - On the Role of 
Ambient Reactive Particles in the Mixing and Afterburn Behind
Explosive Blast Waves; {\sl Combust.\ Sci.\ and Tech.\ } 182, 186–214 (2010).

\bibitem{Ben03} K. Benkiewicz and K. Hayashi -
Two-Dimensional Numerical Simulations of Multi-Headed Detonations in 
Oxygen-Aluminum Mixtures Using an Adaptive Mesh Refinement;
{\sl Shock Waves }, 12, 5, 385-402 (2003).

\bibitem{Bor92} J.P. Boris, F.F. Grinstein, E.S. Oran, and R.J. Kolbe -
New Insights Into Large Eddy Simulation;
{\sl Fluid Dynamics Research }10, 199-228 (1992).

\bibitem{Bur15}
J.A. Burns and C.N. Rautenberg - The Infinite-Dimensional Optimal 
Filtering Problem with Mobile and Stationary Sensor Networks;
{\sl Numerical Functional Analysis and Optimization }36, 2, 181-224 (2015).
https://doi.org/10.1080/01630563.2014.970647

\bibitem{Cam04} F. Camelli and R. L\"ohner - Assessing Maximum Possible Damage
for Contaminant Release Events; {\sl Engineering Computations } 21, 7,
748-760 (2004).

\bibitem{Cam04a} F. Camelli, R. L\"ohner, W.C. Sandberg and R. Ramamurti -
VLES Study of Ship Stack Gas Dynamics; {\sl AIAA}-04-0072 (2004).

\bibitem{Cam06} F. Camelli and R. L\"ohner - VLES Study of Flow 
and Dispersion Patterns in Heterogeneous Urban Areas;
{\sl AIAA}-06-1419 (2006).

\bibitem{Cam11} F. Camelli, J. Lien, D. Dayong, D. W. Wong, 
M. Rice, R. L\"ohner and C. Yang - Generating Seamless Surfaces for 
Transport and Dispersion Modeling in GIS; submitted to
{\sl GeoInformatica } 16, 2, 207-327 (2012).

\bibitem{Ceb01} J.R. Cebral and R. L\"ohner - From Medical Images to
Anatomically Accurate Finite Element Grids;
{\sl Int.\ J.\ Num.\ Meth.\ Eng.\ }51, 985-1008 (2001).

\bibitem{Ceb05} J.R. Cebral and R. L\"ohner - Efficient Simulation of Blood
Flow Past Complex Endovascular Devices Using an Adaptive Embedding
Technique; {\sl IEEE Transactions on Medical Imaging }24, 4, 468-476 (2005).

\bibitem{Cha09} C. Chao, M.P. Wan, L. Morawska, G. Johnson, R. Graham, 
Z. Ristovski M. Hargreaves, K. Mengersen, L. Kerrie C. Steve, 
Y. Li, X. Xie and S. Katoshevski - Characterization of
Expiration Air Jets and Droplet Size Distributions Immediately at the Mouth
Opening; {\sl J.\ of Aerosol Science }40, 2, 122-133 (2009).

\bibitem{Cli78} R. Clift, J.R. Grace and M.E. Weber -
{\sl Bubbles, Drops and Particles}; Academic Press, New York (1978).

\bibitem{Cor10a} A. Corrigan, F. Camelli and R. L\"ohner -
Porting Of An Edge-Based CFD Solver to GPUs;
{\sl AIAA}-10-0523 (2010).

\bibitem{Cor10b} A. Corrigan, F. Camelli, R. L\"ohner and F. Mut - 
Porting of FEFLO to GPUs; {\sl Proc.\ ECCOMAS CFD 2010 Conf.}
Lisbon, Portugal, June 14-17 (2010).

\bibitem{Cor11a} A. Corrigan, F.F. Camelli, R. L\"ohner and J. Wallin -
Running Unstructured Grid Based CFD Solvers on Modern Graphics
Hardware; {\sl Int.\ J.\ Num.\ Meth.\ Fluids }66, 221-229 (2011).

\bibitem{Cor11b} A. Corrigan and R. L\"ohner - Porting of FEFLO to
Multi-GPU Clusters; {\sl AIAA}-11-0948 (2011).

\bibitem{Dee06} N.G. Deen, M. v.Sint Annaland and J.A.M. Kuipers -
Direct Numerical Simulation of Particle Mixing in Dispersed 
Gas-Liquid-Solid Flows Using a Combined Volume of Fluid and Discrete
Particle Approach; {\sl Proc. Fifth Int.\ Conf.\ on CFD in the 
Process Industries}, CSIRO, Melbourne, Australia, 13-15 December (2006). 

\bibitem{Die20} L. Dietz, P.F. Horve, D.A. Coil, M. Fretz,
J.A. Eisen  and L. van den Wymelenberg - 2019
Novel Coronavirus (COVID-19) Pandemic: Built
Environment Considerations to Reduce
Transmission; {\sl mSystems }5:e00245-20 (2020). 
doi:10.1128/mSystems.00245-20.

\bibitem{Fab08}
P. Fabian, J.J. McDevitt, W.H. Dehaan, R.O.P. Fung, B.J. Cowling, 
K.H. Chan, et al.\ - Influenza Virus in Human Exhaled Breath: An 
Observational Study; {\sl PLoS ONE }3:e2691 (2008).

\bibitem{Fra97} D.R. Franz, P.B. Jahrling, A.M. Friedlander, et al. -
Clinical Recognition and Management of Patients Exposed to
Biological Warfare Agents; {\sl JAMA }278:399e411 (1997).

\bibitem{Fur99} C. Fureby and F. Grinstein - Monotonically Integrated
Large Eddy Simulation of Free Shear Flows; {\sl AIAA J.\ }37, 5,
544-556 (1999).

\bibitem{Gri02} F.F. Grinstein and C. Fureby - Recent Progress on MILES
for High-Reynolds-Number Flows; {\sl J.\ Fluids Engineering }124,
848-861 (2002).

\bibitem{Gup09} J.K. Gupta, C-H. Lin and Q. Chen -
Flow Dynamics and Characterization of a Cough; 
{\sl Indoor Air }19, 517–525 (2009).

\bibitem{Gup10} J.K. Gupta, C-H. Lin and Q. Chen -
Characterizing Exhaled Airflow from Breathing and Talking; 
{\sl Indoor Air } 20, 31-39 (2010).

\bibitem{Gup11a} J.K. Gupta, C-H. Lin and Q. Chen -
Inhalation of Expiratory Droplets in Aircraft Cabins;
{\sl Indoor Air }21, 341-350 (2011). 
doi:10.1111/j.1600-0668.2011.00709.x

\bibitem{Gup11b} J.K. Gupta, C-H. Lin and Q. Chen -
Transport of Expiratory Droplets in an Aircraft Cabin;
{\sl Indoor Air } 21, 3-11 (2011).

\bibitem{Hal12} S.K. Halloran, A.S. Wexler and W.D. Ristenpart -
A Comprehensive Breath Plume Model for Disease Transmission via Expiratory 
Aerosols; {\sl PLoS ONE }7(5):e37088 (2012). 
https://doi.org/10.1371/journal.pone.0037088

\bibitem{Hin17} M. Hinterm\"uller, C.N. Rautenberg, M. Mohammadi
and M. Kanitsar - Optimal Sensor Placement: A Robust Approach;
{\sl SIAM J.\ Control Optim.\ }55(6), 3609-3639 (2017).
https://doi.org/10.1137/16M1088867

\bibitem{Hob10} J. Hoberock and N. Bell - Thrust: Parallel Template 
Library, Version 1.3 (2010).

\bibitem{Ide19} S.R. Idelsohn, N. Nigro, A. Larreteguy, J.M. Gimenez 
and P. Ryshakov - A Pseudo-DNS Method for the Simulation of 
Incompressible Fluid Flows with Instabilities at Different Scales;
{\sl  Int. J. Comp.\ Particle Mechanics} (2019).
https://doi.org/10.1007/s40571-019-00264-x

\bibitem{Ip07} M. Ip, J.W. Tang, D.S.C. Hui, A.L.N. Wong, M.T.V. Chan, 
G.M. Joynt, A.T.P. So, S.D. Hall, P.K.S. Chan and J.J.Y. Sung -
Airflow and Droplet Spreading Around Oxygen Masks: A Simulation Model
for Infection Control Research;
{\sl AJIC } 35, 10, 684-689 (2007).

\bibitem{Jam81}
A. Jameson, W. Schmidt and E. Turkel - Numerical Solution
of the Euler Equations by Finite Volume Methods using Runge-Kutta
Time-Stepping Schemes; {\sl AIAA}-81-1259 (1981).

\bibitem{Joh11} G.R. Johnson, L. Morawska, Z.D. Ristovski,
M. Hargreaves, K. Mengersen,
C.Y.H. Chao, M.P. Wan, Y. Li , X. Xie, D. Katoshevski, S. Corbette -
Modality of Human Expired Aerosol Size Distributions
{\sl J.\ of Aerosol Science }42, 839-851 (2011).

\bibitem{Kam20} G. Kampf, D. Todt, S. Pfaender, E. Steinmann -
Persistence of Coronaviruses on Inanimate Surfaces and Their Inactivation 
With Biocidal Agents;
{\sl J.\ of Hospital Infection} 104, 3, 246-251, March 01 (2020).
https://doi.org/10.1016/j.jhin.2020.01.022

\bibitem{Kim08} C.K. Kim, J.G. Moon, J.S. Hwang, M.C. Lai and K.S. Im -
Afterburning of TNT Explosive Products in Air With Aluminum Particles;
{\sl AIAA}-2008-1029 (2008).

\bibitem{Kim20} Y.-I. Kim et al. - 
Infection and Rapid Transmission of SARS-CoV-2 in Ferrets;
{\sl Cell Host and Microbe }27, 1-6 (2020).
https://doi.org/10.1016/j.chom.2020.03.023


\bibitem{Li07} Y. Li et al. - Role of Ventilation in Airborne Transmission 
of Infactious Agents in the Built Environment - A Multidisciplinary 
Systematic Review; {\sl Indoor Air }17, 2-18 (2007).

\bibitem{Lin10}
W.G. Lindsley, F.M. Blachere, R.E. Thewlis, A. Vishnu, K.A. Davis, 
G. Cao, et al.\ - Measurements of Airborne Influenza Virus in Aerosol 
Particles from Human Coughs; {\sl PLoS ONE }5:e15100 (2010).

\bibitem{Lin12}
W.G. Lindsley, T.A. Pearce, J.B. Hudnall, K.A. Davis, S.M. Davis, 
M.A. Fisher, et al.\ - Quantity and Size Distribution of Cough-Generated 
Aerosol Particles Produced by Influenza Patients During and After 
Illness; {\sl J.\ Occup.\ Environ.\ Hyg.\ } 9, 443-9. (2012).

\bibitem{Liu09} J. Liu, K. Kailasanath, R. Ramamurti, D. Munday,
E. Gutmark and R. L\"ohner - Large-Eddy Simulations of a Supersonic
Jet and Its Near-Field Acoustic Properties;
{\sl AIAA J.\ }47, 8, 1849-1864 (2009).

\bibitem{Loh87} R. L\"ohner, K. Morgan, J. Peraire and M. Vahdati - 
Finite Element Flux-Corrected Transport (FEM-FCT) for the Euler 
and Navier-Stokes Equations; {\sl Int.\ J.\ Num.\ Meth.\ Fluids }7,
1093-1109 (1987).

\bibitem{Loh90} R. L\"ohner and J. Ambrosiano - A Vectorized Particle Tracer
for Unstructured Grids; {\sl J.\ Comp.\ Phys.\ }91, 1, 22-31 (1990).

\bibitem{Loh95} R. L\"ohner and R. Ramamurti - A Load Balancing Algorithm for
Unstructured Grids; {\sl Comp.\ Fluid Dyn.\ }5, 39-58 (1995).

\bibitem{Loh98} R. L\"ohner - Renumbering Strategies for Unstructured-Grid
Solvers Operating on Shared-Memory, Cache-Based Parallel Machines;
{\sl Comp.\ Meth.\ Appl.\ Mech.\ Eng.\ }163, 95-109 (1998).

\bibitem{Loh98c} R. L\"ohner, C. Yang and E. O\~nate -
Viscous Free Surface Hydrodynamics Using Unstructured Grids;
{\sl Proc.\ 22nd Symp.\ Naval Hydrodynamics}, Washington, D.C.,
August (1998).

\bibitem{Loh01} R. L\"ohner, Chi Yang, J. Cebral, O. Soto, F. Camelli,
J.D. Baum, H. Luo, E. Mestreau, D. Sharov, R. Ramamurti, W. Sandberg
and Ch. Oh - Advances in FEFLO; {\sl AIAA}-01-0592 (2001).

\bibitem{Loh02} R. L\"ohner and M. Galle - Minimization of Indirect Addressing
for Edge-Based Field Solvers; {\sl Comm.\ Num.\ Meth.\ Eng.\ }18, 335-343
(2002).

\bibitem{Loh04} R. L\"ohner - Multistage Explicit Advective Prediction for
Projection-Type Incompressible Flow Solvers;
{\sl J.\ Comp.\ Phys.\ }195, 143-152 (2004).

\bibitem{Loh04a} R. L\"ohner, J.D. Baum and D. Rice -
Comparison of Coarse and Fine Mesh 3-D Euler Predictions
for Blast Loads on Generic Building Configurations;
{\sl Proc. MABS-18 Conf.}, Bad Reichenhall, Germany, September (2004).

\bibitem{Loh05} R. L\"ohner and F. Camelli - Optimal Placement of
Sensors for Contaminant Detection Based on Detailed 3-D CFD
Simulations; {\sl Engineering Computations }22, 3, 260-273 (2005).

\bibitem{Loh06} R. L\"ohner, Chi Yang, J.R. Cebral, F. Camelli, O. Soto and
J. Waltz - Improving the Speed and Accuracy of Projection-Type Incompressible
Flow Solvers; {\sl Comp.\ Meth.\ Appl.\ Mech.\ Eng.\ }195, 23-24, 3087-3109
(2006).

\bibitem{Loh06a} R. L\"ohner, Chi Yang and E. O\~nate - On the Simulation
of Flows with Violent Free Surface Motion;
{\sl Comp.\ Meth.\ Appl.\ Mech.\ Eng.\ }195, 5597-5620 (2006).

\bibitem{Loh07b} R. L\"ohner, Chi Yang and E. O\~nate -
Simulation of Flows With Violent Free Surface Motion and Moving
Objects Using Unstructured Grids;
{\sl Int.\ J.\ Num.\ Meth.\ Fluids } 53, 1315-1338 (2007).

\bibitem{Loh08} R. L\"ohner - {\sl Applied CFD Techniques, 
Second Edition}; J. Wiley \& Sons (2008).

\bibitem{Loh08a} R. L\"ohner, H. Luo, J.D. Baum and D. Rice -
Improvements in Speed for Explicit, Transient Compressible Flow
Solvers;
{\sl Int.\ J.\ Num.\ Meth.\ Fluids } 56, 12, 2229-2244 (2008).

\bibitem{Loh08b} R. L\"ohner, J.R. Cebral, F.F. Camelli,
S. Appanaboyina, J.D. Baum, E.L. Mestreau and O. Soto - Adaptive 
Embedded and Immersed Unstructured Grid Techniques;
{\sl Comp.\ Meth.\ Appl.\ Mech.\ Eng.\ }197, 2173-2197 (2008).

\bibitem{Loh11a} R. L\"ohner, F. Mut and F.F. Camelli - Timings OF FEFLO
on the SGI-ICE Machines; {\sl AIAA}-11-1064 (2011).

\bibitem{Loh11b} R. L\"ohner and A. Corrigan - Semi-Automatic Porting 
if a General Fortran CFD Code to GPUs: The Difficult Modules; 
{\sl AIAA}-11-3219 (2011).

\bibitem{Loh11c} R. L\"ohner, F. Mut, J.R. Cebral, R. Aubry and
G. Houzeaux; Deflated Preconditioned Conjugate Gradient Solvers
for the Pressure-Poisson Equation:
Extensions and Improvements; {\sl Int.\ J.\ Num.\ Meth.\ Eng.\ }87, 1-5,
2-14 (2011).

\bibitem{Loh12} R. L\"ohner - F2GPU - A General Fortran to GPU
Translator; {\sl Proc. NVIDIA GTC Conf.\ }, San Jose, CA, May (2012).

\bibitem{Loh14} R. L\"ohner, F. Camelli, J.D. Baum, F. Togashi and
O. Soto - On Mesh-Particle Techniques; {\sl Comp.\ Part.\ Mech.\ }1,
199-209 (2014).

\bibitem{Loh16} R. L\"ohner and F. Camelli -
Tightly Coupled Computational Fluid and Crowd Dynamics; pp.
505-509 in
{\sl Proc. Pedestrian and Evacuation Dynamics 2016 (PED~2016)},
(W. Song, J. Ma and L. Fu eds.), University of Science and
Technology Press, Hefei, China, Oct 17-21 (2016).

\bibitem{Lou67} R.G. Loudon and R.M. Roberts - Droplet Expulsion from the 
Respiratory Tract; {\sl Am.\ Rev.\ Respir.\ Dis.\ } 95, 3, 435–442 (1967).

\bibitem{Luo94} H. Luo, J.D. Baum and R. L\"ohner - Edge-Based 
Finite Element Scheme for the Euler Equations; 
{\sl AIAA J.\ }32, 6, 1183-1190 (1994).

\bibitem{Luo94a} H. Luo, J.D. Baum, R. L\"ohner and J. Cabello - Implicit
Finite Element Schemes and Boundary Conditions for Compressible Flows
on Unstructured Grids; {\sl AIAA}-94-0816 (1994).

\bibitem{Luo99} H. Luo, J.D. Baum and R. L\"ohner - An Accurate, Fast,
Matrix-Free Implicit Method for Computing Unsteady Flows on Unstructured
Grids; {\sl AIAA}-99-0937 (1999).

\bibitem{Luo00} H. Luo, D. Sharov, J.D. Baum and R. L\"ohner -
A Class of Matrix-free Implicit Methods for Compressible Flows
on Unstructured Grids; {\sl First International Conference
on Computational Fluid Dynamics}, Kyoto, Japan, July 10-14 (2000).

\bibitem{Luo01} H. Luo, J.D. Baum and R. L\"ohner - A Fast, Matrix-Free
Implicit Method for Computing Low Mach Number Flows on Unstructured Grids;
{\sl Int.\ J.\  CFD }14, 133-157 (2001).

\bibitem{Mer10} D. Merrill and A. Grimshaw -
Revisiting Sorting for GPGPU Stream Architectures;
{\sl UVA CS Rep. CS2010-03}, Charlottesville, VA (2010).

\bibitem{Mil13}
D.K. Milton, M.P. Fabian, B.J. Cowling, M.L. Grantham, J.J. McDevitt -
Influenza Virus Aerosols in Human Exhaled Breath: Particle Size, 
Culturability, and Effect of Surgical Masks; {\sl PLoS Pathog.\ } 9:e1003205
(2013).

\bibitem{Nvi10_32} NVIDIA Corporation. NVIDIA CUDA 3.2 Programming 
Guide (2010).

\bibitem{Per87} J. Peraire, M. Vahdati, K. Morgan and O.C. Zienkiewicz -
Adaptive Remeshing for Compressible Flow Computations;
{\sl J.\ Comp.\ Phys. }72, 449-466 (1987).

\bibitem{Pet09} P. Peterson - F2PY: Tool for Connecting Fortran 
and Python Programs; {\sl Int. J. Computational Science
and Engineering }4, 296-305 (2009).

\bibitem{Ram93} R. Ramamurti and R. L\"ohner - Simulation of Flow Past Complex
Geometries Using a Parallel Implicit Incompressible Flow Solver;
pp. 1049,1050 in {\sl Proc.\ 11th AIAA CFD Conf.\ }, Orlando, FL,
July (1993).

\bibitem{Ram96} R. Ramamurti and R. L\"ohner - A Parallel Implicit
Incompressible Flow Solver Using Unstructured Meshes;
{\sl Computers and Fluids } 5, 119-132 (1996).

\bibitem{Ram99} R. Ramamurti, W.C. Sandberg and R. L\"ohner - Computation of
Unsteady Flow Past Deforming Geometries; {\sl Int.\ J.\ Comp.\ Fluid Dyn.\ },
83-99 (1999).

\bibitem{Ric08} D.L. Rice, J.D. Baum, F. Togashi, R. L\"ohner 
and A. Amini - First-Principles Blast Diffraction Simulations on a 
Notebook: Accuracy, Resolution and Turn-Around Issues;
{\sl Proc. MABS-20 Conf.}, Oslo, Norway, September (2008).

\bibitem{Sch79} H. Schlichting - {\sl Boundary Layer Theory}; 
McGraw-Hill (1979).

\bibitem{Sha00} D. Sharov, H. Luo, J.D. Baum and R. L\"ohner - Implementation
of Untructured Grid GMRES+LU-SGS Method on Shared-Memory, Cache-Based
Parallel Computers; {\sl AIAA}-00-0927 (2000).

\bibitem{Stu10} A. St\"uck, F. Camelli and R. L\"ohner -
Adjoint-Based Design of Shock Mitigation Devices;
{\sl Int.\ J.\ Num.\ Meth.\ Fluids } 64, 443-472 (2010).

\bibitem{Tan06} J.W. Tang, Y. Li, I. Eames, P.K.S. Chan and G.L. Ridgway
Factors Involved in the Aerosol Transmission of Infection and 
Control of Ventilation in Healthcare Premises;
{\sl J.\ of Hospital Infection }64, 100-114 (2006).

\bibitem{Tan11} J.W. Tang, C.J. Noakes, P.V. Nielsen, I. Eames, 
A. Nicolle, Y. Li and
G.S. Settles - Observing and Quantifying Airflows in the Infection 
Control of Aerosol- and Airborne-Transmitted Diseases: An Overview of 
Approaches; {\sl J.\ of Hospital Infection }77 213-222 (2011).

\bibitem{Tan12} J.W. Tang, A.D. Nicolle, J. Pantelic, G.C. Koh, L. Wang,
M. Amin, C.A. Klettner, D.K.W. Cheong, C. Sekhar and K.W. Tham -
Airflow Dynamics of Coughing in Healthy Human Volunteers by Shadowgraph
Imaging: An Aid to Aerosol Infection Control;
{\sl PLoS ONE }7, 4: e34818 (2012).
doi:10.1371/journal.pone.0034818

\bibitem{Tan13} J.W. Tang, A.D. Nicolle, C.A. Klettner, J. Pantelic,
L. Wang, A. Bin Suhaimi, A.Y.L. Tan, G.W.X. Ong, R. Su, C. Sekhar,
D.K.W. Cheong and K.W. Tham - Airflow Dynamics of Human Jets: Sneezing and
Breathing - Potential Sources of Infectious Aerosols;
{\sl PLoS ONE }8, 4: e59970 (2013).
doi:10.1371/journal.pone.0059970

\bibitem{Teu10} P.F.M. Teunis, N. Brienen, M.E.E. Kretzschmar -
High Infectivity and Pathogenicity of Influenza A Virus Via Aerosol and
Droplet Transmission; {\sl Epidemics }2, 215–222 (2010).

\bibitem{Til08} R. Tilch, A. Tabbal, M. Zhu, F. Decker and R. L\"ohner -
Combination of Body-Fitted and Embedded Grids for External Vehicle
Aerodynamics; {\sl Engineering Computations }25, 1, 28-41 (2008).

\bibitem{To20} K. K.-W. To et al. - 
Temporal Profiles of Viral Load in Posterior Oropharyngeal
Saliva Samples and Serum Antibody Responses During
Infection by SARS-CoV-2: An Observational Cohort Study;
{\sl Lancet Infect.\ Dis.\ } (online) (2020).
https://doi.org/10.1016/S1473-3099(20)30196-1

\bibitem{Tog09} F. Togashi, J.D. Baum, E. Mestreau, R. L\"ohner,
and D. Sunshine;
Numerical Modeling of Long-Duration Blast Wave Evolution in
Confined Facilities; {\sl AIAA}-09-1531 (2009).

\bibitem{vDo20} N. van Doremalen, T. Bushmaker, D.H. Morris,
M.G. Holbrook, A. Gamble, B.N. Williamson, A. Tamin, J.L. Harcourt,
N.J. Thornburg, S.I. Gerber, J.O. Lloyd-Smith, E. de Wit, V.J. Munster -
Aerosol and Surface Stability of SARS-CoV-2 as Compared with SARS-CoV-1;
{\sl The New England Journal of Medicine }382, 16, April 16 (2020).
DOI: 10.1056/NEJMc2004973

\bibitem{Vre06} A.W. Vreman, B.J. Geurts, N.G. Deen and J.A.M. Kuipers -
Large-Eddy Simulation of a Particle Laden Turbulent Channel Flow;
{\sl Proc. Direct and Large-Eddy Simulation V},
ERCOFTAC Series Volume 9, pp 271-278  (2004).

\bibitem{Wei16} J. Wei, Y. Li - Airborne Spread of Infectious Agents 
in the Indoor 
Environment; {\sl American J.\ of Infection Control }44, S102-S108 (2016).
http://dx.doi.org/10.1016/j.ajic.2016.06.003

\bibitem{Xie07} X. Xie, Y. Li, A.T.Y. Chwang, P.L. Ho, W.H. Seto -
How Far Droplets Can Move in Indoor Environments - Revisiting
the wells Evaporation-Falling Curve;
{\sl Indoor Air }17, 211-225 (2007). 
doi:10.1111/j.1600-0668.2006.00469.x

\bibitem{Zha07} T. Zhang, Q. Chen and C.-H. Lin - Optimal Sensor 
Placement for Airborne Contaminant Detection in an Aircraft Cabin;
{\sl HVAC\&R Research }13, 5, 683-696 (2007).

\bibitem{Zha17} Y. Zhang, G. Feng, Z. Kang, Y. Bi and Y. Cai - 
Numerical Simulation of Coughed Droplets in Conference Room;
{\sl 10th International Symposium on Heating, Ventilation and Air 
Conditioning, ISHVAC2017}, October, 19-22 Jinan, China (2017),
{\sl Procedia Engineering }205, 302–308 (2017).

\end{thebibliography}

\end{document}